\documentclass[fleqn,usenatbib]{mnras}

\usepackage[T1]{fontenc}

\DeclareRobustCommand{\VAN}[3]{#2}
\let\VANthebibliography\thebibliography
\def\thebibliography{\DeclareRobustCommand{\VAN}[3]{##3}\VANthebibliography}

\usepackage{graphicx}	
\usepackage{amsmath}	
\usepackage{amssymb}	
\usepackage{standalone}
\usepackage{tikz}
\usepackage{tikz-3dplot} 
\usepackage{bm} 
\usepackage{caption}
\usepackage{subcaption}
\usepackage{color,soul}

\newcommand{\vth}{v_\text{th}}
\newcommand{\M}{M_\mathrm{BH}}
\newcommand{\fom}{\mathcal{M}}
\newcommand{\msun}{\mathrm{M}_\odot}
\newcommand{\dd}{\mathrm{d}}

\newcommand{\qo}{\citetalias{quera-bofarull_qwind_2020}}

\newcommand{\sigmat}{\sigma_{\text{\tiny T}}}
\newcommand{\sigmae}{\kappa_{\text{\tiny e}}}
\newcommand{\sigmax}{\sigma_\text{\tiny X}}
\newcommand{\sigmasb}{\sigma_\text{\tiny SB}}
\newcommand{\fuv}{f_\text{\tiny UV}}
\newcommand{\fnt}{f_\text{\tiny NT}}
\newcommand{\tauuv}{\tau_\text{\tiny UV}}
\newcommand{\taux}{\tau_\text{\tiny X}}
\newcommand{\logt}{\log_{10}}
\newcommand{\rd}{R_\mathrm{d}}
\newcommand{\rin}{R_\mathrm{in}}
\newcommand{\pd}{\phi_\mathrm{d}}
\newcommand{\rg}{R_\mathrm{g}}
\newcommand{\isco}{R_\mathrm{isco}}
\newcommand{\phid}{\phi_\mathrm{d}}
\newcommand{\qw}{\textsc{Qwind}}

\title[Qwind3]{Qwind3: UV line-driven accretion disc wind models for AGN feedback}

\author[Arnau Quera-Bofarull et al.]{Arnau Quera-Bofarull,$^{1,2}$\thanks{E-mail: arnau.quera-bofarull@durham.ac.uk}
Chris Done$^{3}$,
Cedric G. Lacey$^{1}$,
Mariko Nomura$^{4,5}$,
Ken Ohsuga$^6$
\\
$^{1}$Institute for Computational Cosmology, Department of Physics, Durham University, South Road, Durham DH1 3LE, UK\\
$^{2}$Institute for Data Science, Durham University, South Road, Durham DH1 3LE, UK\\
$^{3}$Centre for Extragalactic Astronomy, Department of Physics, Durham University, South Road, Durham DH1 3LE, UK\\
$^{4}$Faculty of Natural Sciences, National Institute of Technology (KOSEN), Kure College, 2-2-11 Agaminami, Kure, Hiroshima 737-8506, Japan\\
$^{5}$Astronomical Institute, Graduate School of Science, Tohoku University, 6-3 Aoba, Aramaki, Aoba-ku, Sendai, Miyagi 980-8578, Japan\\
$^{6}$Center for Computational Sciences, University of Tsukuba, Ten-nodai, 1-1-1 Tsukuba, Ibaraki 305-8577, Japan\\
}

\date{Accepted XXX. Received YYY; in original form ZZZ}

\pubyear{2021}

\usepackage{newtxtext,newtxmath}

\begin{document}
\label{firstpage}
\pagerange{\pageref{firstpage}--\pageref{lastpage}}
\maketitle

\begin{abstract}
    The ultraviolet (UV) bright accretion disc in active galactic nuclei (AGN) should give rise to line driving, producing a powerful wind which may play an important role in AGN feedback as well as in producing structures like the broad line region. However, coupled radiation-hydrodynamics codes are complex and expensive, so we calculate the winds instead using a non-hydrodynamical approach (the \textsc{Qwind} framework). The original \textsc{Qwind} model assumed the initial conditions in the wind, and had only simple radiation transport. Here, we present an improved version which derives the wind initial conditions and has significantly improved ray-tracing to calculate the wind absorption self consistently given the extended nature of the UV emission. We also correct the radiation flux for relativistic effects, and assess the impact of this on the wind velocity. These changes mean the model is more physical, so its predictions are more robust. We find that, even when accounting for relativistic effects, winds can regularly achieve velocities $\simeq (0.1-0.5)\; c$, and carry mass loss rates which can be up to 30\% of the accreted mass for black hole masses of $10^{7-9}\msun$, and mass accretion rates of 50\% of the Eddington rate. Overall, the wind power scales as a power law with the black hole mass accretion rate, unlike the weaker scaling generally assumed in current cosmological simulations that include AGN feedback. The updated code, \textsc{Qwind3}, is publicly available in GitHub\footnotemark.

\end{abstract}

\begin{keywords}
keyword1 -- keyword2 -- keyword3
\end{keywords}

\footnotetext{https://github.com/arnauqb/Qwind.jl}

\defcitealias{laor_line-driven_2014}{LD14}
\defcitealias{pereyra_steady_2004}{PE04}
\defcitealias{pereyra_steady_2006}{PE06}
\defcitealias{castor_radiation-driven_1975}{CAK}
\defcitealias{quera-bofarull_qwind_2020}{Q20}
\defcitealias{stevens_x-ray_1990}{SK90}
\defcitealias{nomura_modeling_2013}{N13}
\defcitealias{proga_dynamics_2000}{P00}
\defcitealias{proga_04}{P04}



\section{Introduction}

AGN feedback is a very important process in shaping the growth of galaxies, but the prescriptions for it that are included in current cosmological simulations are generally highly simplified and 
not based on any deeper understanding of the physical processes involved. Jets from AGN are poorly understood, but some types of AGN winds can be calculated ab initio from the fundamental parameters of black hole mass, mass accretion rate and spin. Observations show the existence of ultra-fast outflows (UFOs) in AGN, likely originating from the accretion disc close to the central supermassive black hole (BH). These outflows can reach velocities of $v$ $\sim (0.03-0.3)\,c$ \citep{weymann_comparisons_1991, pounds_evidence_2003, pounds_high-velocity_2003, reeves_compton-thick_2009, crenshaw_feedback_2012, tombesi_evidence_2010, fiore_agn_2017}. UV line driving is a mechanism which is especially likely to be present in luminous AGN, with their accretion disc spectrum peaking in the UV, where there are multiple strong atomic transitions in low ionisation material. These transitions absorb the photon momentum, producing the strong winds seen from similar temperature material in O star photospheres \citep{howarth_stellar_1989}, which were first extensively studied by \cite{castor_radiation-driven_1975} (hereafter  \citetalias{castor_radiation-driven_1975}) and \cite{abbott_theory_1980}. In the context of AGN, the study of UV line-driven winds started with analytical studies \citep{murray_accretion_1995}, and continued with the use of radiation-hydrodynamic simulations \citep{proga_radiation-driven_1998, nomura_radiation_2016}. However, the computational complexity of the radiation-hydrodynamics codes prevents us from efficiently exploring the input parameter space. Even more importantly, the complexity of these codes can obscure the effect of some of the underlying assumptions, e.g. the lack of scattered emission, on setting the radiation environment \citep{higginbottom_line-driven_2014}, or the effect of wind mass loss on the net accretion rate and hence the disc emission \citep{nomura_line-driven_2020}. 

To circumvent this, we build on the pioneering approach of \textsc{Qwind} \citep{risaliti_non-hydrodynamical_2010} in developing a non-hydrodynamic code. This calculates ballistic trajectories, ignoring pressure forces (which should be negligible in a supersonic flow) but including gravity and radiation forces, to obtain the streamlines, making the computer code much faster, and simpler, so that it can be used to explore the parameter space much more fully. In \cite{quera-bofarull_qwind_2020} (hereafter \citetalias{quera-bofarull_qwind_2020}) we released a modern version of this code (\textsc{Qwind2}), but this was still based on some underlying, arbitrary parameter choices, including for the launching of the wind from the accretion disk, and used simplified radiation transport. 
Here, we aim to significantly improve the predictive power of the \textsc{Qwind} code. We present a model to derive the initial conditions of the wind, a radiative transfer algorithm that takes into account the wind geometry and density structure, and we include special relativistic corrections to our calculations of the radiation force on the wind. We then use this new model, which we refer as \textsc{Qwind3}, to study the dependence of the mass loss rate and kinetic power of the wind on the BH mass and mass accretion rate. These results can form the basis for a physical prescription for AGN feedback that can be used in cosmological simulations to explore the coeval growth of galaxies and their central black holes across cosmic time. 

\section{Review of Qwind}

The \textsc{Qwind} code of Q20 is based on the approach of \cite{risaliti_non-hydrodynamical_2010}, which calculates ballistic trajectories of gas blobs launched from the accretion disc. These blobs are subject to two forces: the gravitational pull of the BH, and the outwards pushing radiation force, which can be decomposed  into an X-ray and a UV component, with the later being dominant. On the one hand, the X-ray photons couple to the accretion disc material via bound-free transitions with outer electrons, ionising the gas. On the other hand, the UV opacity is greatly enhanced when the material is not over-ionised, since the UV photons can then excite electrons to higher states while transferring their momentum to the gas in the process. If sufficient momentum is transferred from the radiation field to the gas, the latter may eventually reach the gravitational escape velocity, creating an outgoing flow. This mechanism for creating a wind is known as UV line-driving. The conditions under which the wind can escape depend on the density and velocity structure of the flow, where part of the material can be shielded from the X-ray radiation while being illuminated by the UV emitting part of the accretion disc. 

\subsection{Radiation force}

Using a cylindrical coordinate system $(R, \phi, z)$, with $r^2 = R^2 + z^2$, let us consider a BH of mass $M_\mathrm{BH}$, located at $r=0$, accreting mass at a rate $\dot M$, and an accretion disc located at the $z=0$ plane. We use the gravitational radius, $R_g = G \M / c^2$, as our natural unit of length. The total luminosity of the system is related to the accreted mass through
\begin{equation}
    L_\text{bol} = \eta \dot M c^2
\end{equation}
where $\eta$ is the radiation efficiency. We set $\eta = 0.057$ throughout this work, as we only consider non-rotating BHs \citep{thorne_disk-accretion_1974}. We frequently refer to  the Eddington fraction $\dot m = \dot M / \dot M_\text{Edd}$, where $\dot M_\text{Edd}$ is the mass accretion rate corresponding to the Eddington luminosity,
\begin{equation}
    \dot M_\text{Edd} = \frac{L_\text{Edd}}{\eta c^2} = \frac{4\pi G \M}{\eta c \sigmae},
\end{equation}
where $\sigmae$ is the electron scattering opacity, related to the electron scattering cross section $\sigmat$ through
\begin{equation}
    \sigmae = \frac{\sigmat}{m_p \, \mu_e},
\end{equation}
where $m_p$ is the proton mass, and $\mu_e$ is the mean molecular weight per electron. We set $\mu_e = 1.17$ corresponding to a fully ionised gas with solar chemical abundance \citep{asplund_chemical_2009}.

The emitted UV radiated power per unit area by a disc patch located at $(\rd, \phid, 0)$ is given by \citep{shakura_black_1973}
\begin{equation}
    \label{eq:radiation_flux}
    \mathcal F_{\rm UV} = \frac{3 G  \M \dot M}{8\pi \rd^3} \fuv(\rd) \fnt(\rd, \isco),
\end{equation}
where $\fnt$ are the Novikov-Thorne relativistic factors \citep{novikov_astrophysics_1973}, and $\fuv$ is the fraction of power in the UV band, which we consider to be (200--3200) \AA. (The total power radiated per unit area $\mathcal F$ is given by setting $\fuv=1$ in the above equation.) The force per unit mass exerted on a gas blob at a position $(R, 0, z)$ due to electron scattering is (see \citetalias{quera-bofarull_qwind_2020})
\begin{equation}
    \label{eq:radiation_acceleration}
        \bmath{a}_\mathrm{rad}^\mathrm{es}(R,z) = \mathcal C \,z\int\int\frac{\fuv \fnt}{\rd^2 \, \Delta^4} e^{-\tauuv} \begin{pmatrix}R-\rd\cos\phid\\ -\rd \sin \pd\\ z \end{pmatrix} \, \dd \rd \dd \phid,
\end{equation}
where 
\begin{equation}
    \mathcal C = \frac{3 G \M \dot M \sigmae}{8 \pi^2 c},
\end{equation}
$\Delta^2 = R^2 + z^2 - 2 R \rd \cos\phid$, and $\tauuv$ is the UV optical depth measured from the disc patch to the gas blob. We note that it is enough to consider the case $\phi=0$ due to the axisymmetry of the system, and, furthermore, the $\phi$ component of the force vanishes upon integration. The total radiation force can be greatly amplified by the contribution from the line opacity, which we parameterise as $\kappa_\mathrm{line} = \fom \,\sigmae$, such that the total radiation opacity is $(1 + \fom) \sigmae$, implying that
\begin{equation}
    \bmath{a}_\mathrm{rad} = (1 + \mathcal M) \; \bmath{a}_\mathrm{rad}^\mathrm{es}.
\end{equation}
The parameter $\mathcal M$ is known as the force multiplier, and we use the same parametrisation as \citetalias{quera-bofarull_qwind_2020} \citep{stevens_x-ray_1990} (hereafter \citetalias{stevens_x-ray_1990}). A limitation of our assumed parametrisation is that we do not take into account the dependence of the force multiplier on the particular spectral energy distribution (SED) of the accretion disc \citep{dannen_photoionization_2019}. Furthermore, the force multiplier is also expected to depend on the metallicity of the gas \citep{nomura_radiation_2021}. A self consistent treatment of the force multiplier with relation to the accretion disc and its chemical composition is left to future work. 

It is useful to consider that, close to the disc's surface, the radiation force is well approximated by considering the radiation force produced by an infinite plane at a temperature equal to the local disc temperature,
\begin{equation}
    \label{eq:radiation_acceleration_approx}
    \bmath{a}_\mathrm{rad, 0}^\mathrm{es}(R) = \frac{3 G\M \dot M \sigmae}{8\pi^2 R^3 c} \fuv \fnt\, e^{-\tauuv}\; \bmath{\hat{z}},
\end{equation}
where $\tauuv$ is calculated along a vertical path. The radiation force is then vertical and almost constant at small heights 
($z \lesssim 0.1 \rg$).
To speed up calculations and minimise numerical errors, we use this expression when $z < 0.01 \rg$.

\subsection{Equations of motion}

The equations of motion of the gas blob trajectories are

\begin{equation}
\label{eq:trajectory_ode}
    \begin{split}
        &\frac{\dd R}{\dd t} &=& \; v_R,\\
        &\frac{\dd z}{\dd t} &= &\; v_z,\\
        &\frac{\dd v_R}{\dd t} &= &\; a^\mathrm{grav}_R + a^\mathrm{rad}_R + \frac{\ell^2}{R^3},\\
        &\frac{\dd v_z}{\dd t} &= &\; a^\mathrm{grav}_z + a^\mathrm{rad}_z,
    \end{split}
\end{equation}
where $\ell$ is the specific angular momentum (assumed constant for a given blob), and $\bmath{a}_\mathrm{grav}$ is the gravitational acceleration,
\begin{equation}
\label{eq:gravity}
\bmath{a}_\mathrm{grav}\,(R,z) = -\frac{G\M}{r^2}\,\begin{pmatrix}R/r \\ 0 \\ z / r\end{pmatrix}.
\end{equation}
We assume that initially the gas blobs are in circular orbits around the BH, so that $\ell = \sqrt{G \M R_0}$, where $R_0$ is the launch radius, and thus the azimuthal velocity component at any point is $v_\phi = \ell / r$.

As in \qo, we ignore contributions from gas pressure (except when calculating the launch velocity and density) as these are negligible compared to the radiation force, especially since we focus our study on the supersonic region of the wind. We assume that the distance between two nearby trajectories at any point, $\Delta r$, is proportional to the distance to the centre $\Delta r \propto r$, so that the surface mass loss rate along a particular streamline is $\dot \Sigma = \dot M_\mathrm{streamline} / (2 \pi \, r_0 \, \Delta r_0)$. This both captures the fact that streamlines are mostly vertical close to the disc, and diverge in a cone-like shape at large distances. The gas blob satisfies the approximate mass conservation equation along its trajectory  (\citetalias{quera-bofarull_qwind_2020}),
\begin{equation}
    \label{eq:mass_conservation}
    \dot M_\mathrm{streamline} = 2 \, \pi \, r \, \Delta r \, \rho \, v,
\end{equation}
where $\rho$ is the density of the wind, related to the number density $n$ through $\rho = \mu \, m_p \, n$, and $\Delta r = (r / r_0) \Delta R_0$. We set the mean molecular weight $\mu$ to $\mu = 0.61$, corresponding to a fully ionised gas with solar abundance \citep{asplund_chemical_2009}. We use \autoref{eq:mass_conservation} to determine the gas density at each point along a trajectory.

\subsection{Improvements to \qw}

In \qo, a series of assumptions are made to facilitate the numerical solution of the presented equations of motion. Furthermore, the initial conditions of the wind are left as free parameters to explore, limiting the predictive power of the model. In this work, we present a series of important improvements to the \textsc{Qwind} code. 
Firstly we calculate $\fuv$ from the disc spectrum rather than have this as a free parameter (\autoref{sec:fuv}).
Secondly, we derive the initial conditions of the wind in \autoref{sec:initial_conditions}, based on the methodology introduced in \citetalias{castor_radiation-driven_1975}, and further developed in \cite{abbott_theory_1982, pereyra_steady_2004, pereyra_further_2005, pereyra_steady_2006}. This removes the wind's initial velocity and density as degrees of freedom of the system. Thirdly, we vastly enhance the treatment of the radiative transfer in the code, reconstructing the wind density and velocity field from the calculated gas trajectories. This allows us to individually trace the light rays coming from the accretion disc and the central X-ray source, correctly accounting for their attenuation. This is explained in detail in \autoref{sec:radiation_transport}. Lastly, we include the relativistic corrections from \cite{luminari_importance_2020} in the calculation of the radiation force, solving the issue of superluminal winds, and we later compare our findings to \cite{luminari_speed_2021}.
Readers interested only in the results should skip to Section \ref{sec:results}. 

The improvement in the modelling of the physical processes comes at the expense of added computational cost. We have ported the \textsc{Qwind} code to the Julia programming language \citep{bezanson_julia_2017}, which is an excellent framework for scientific computing given its state of the art performance, and ease of use. The new code is made available to the community under the GPLv3 license on GitHub\footnote{https://github.com/arnauqb/Qwind.jl}.

\section{Radial dependence of \texorpdfstring{$\fuv$}{fuv}}
\label{sec:fuv}

\begin{figure}
    \centering
    \includegraphics[width=\columnwidth]{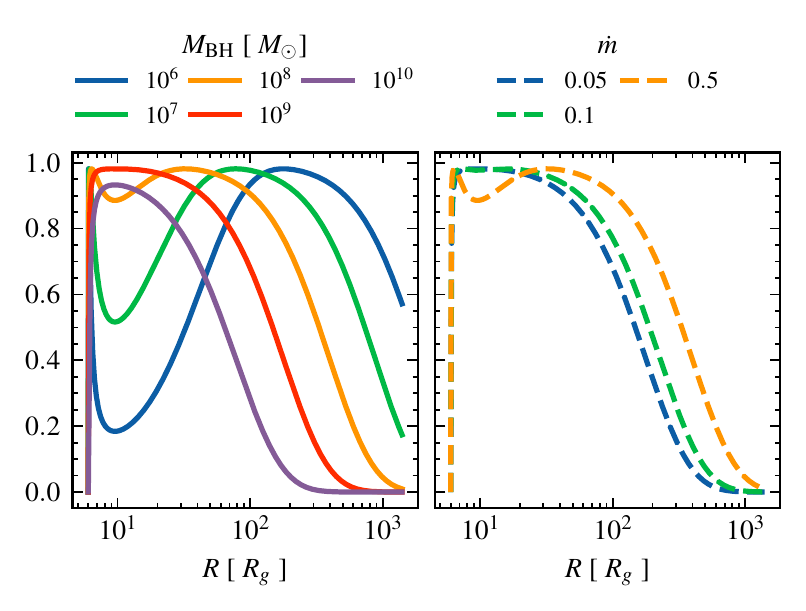}
    \caption{UV fraction as a function of disc radius. Left panel: dependence on $\M$ for fixed $\dot m =0.5$. Right panel: dependence on $\dot m$ for fixed $\M = 10^8\msun$.}
    \label{fig:uv_fractions}
\end{figure}

We first address the validity of assuming a constant emitted UV fraction with radius. We can calculate its radial dependence using
\begin{equation}
    \fuv(\rd) = \frac{\int_{E_1}^{E_2} B(E, T(\rd))\, \dd E}{\int_{0}^{\infty} B(E, T(\rd))\, \dd E},
\end{equation}
where $B(E,T)$ is the blackbody spectral radiance, $E_1 = 0.0038$ keV and $E_2=0.06$ keV (the standard definition of the UV transition band: (3200-200) \AA), and $T^4(\rd)=\mathcal{F} / \sigmasb$ (where $\mathcal F$ is defined in \autoref{eq:radiation_flux}). In \autoref{fig:uv_fractions}, we plot the UV fraction as a function of radius for different $\M$ and $\dot m$. The disc temperature is related to the total flux (given by setting $\fuv=1$ in \autoref{eq:radiation_flux}) so $T^4\propto (M\dot{M}\fnt/R^3)\propto \dot{m}/M (R/\rg)^3$.
Thus the disc temperature increases with decreasing $R/\rg$, and for the fiducial case of $\M=10^8 \msun$ this leads to the majority of the disc emission in the UV coming from $R/\rg\le 100$ (left panel of \autoref{fig:uv_fractions}: orange line).
However, the increase in disc temperature at a given $R/\rg$ for decreasing mass means that the same $\dot{m}$ for $\M=10^6 \msun$ gives a UV flux which peaks at $R/\rg>100$, as the inner regions are too hot to emit within the defined UV bandpass (left panel of \autoref{fig:uv_fractions}: blue line). Conversely, for the highest BH masses of $\M=10^{10}\msun$ the disc is so cool that it emits UV only very close to the innermost stable circular orbit (left panel of \autoref{fig:uv_fractions}: purple line). The universal upturn at $R=10\rg$ is caused by the the temperature sharply decreasing at the inner edge of the accretion disc due to the viscuous torque dropping to zero there.

Similarly, the right panel of \autoref{fig:uv_fractions} shows the impact of changing $\dot{m}$ for the fiducial mass of $10^8\msun$. The dashed orange line shows the case $\dot{m}=0.5$, as before, 
and the disc temperature decreases with decreasing $\dot{m}$ to $0.1$ (dashed green line) and $0.05$ (dashed blue line), reducing the radial extent of the UV-emitting zone. 

We note that the assumption used in \citetalias{quera-bofarull_qwind_2020} (and many other UV line driven disc wind codes) of $\fuv$ being a constant value is poor overall, highlighting the importance of including the radial dependence of the UV flux.

\section{Initial conditions}
\label{sec:initial_conditions}

As initial conditions, we determine the density and velocity at the base of the wind following the \citetalias{castor_radiation-driven_1975} formalism. Let us consider a wind originating from the top of an accretion disc. At low heights, a gas blob is mostly irradiated by the local region of the disc that is just below it. Since this local disc area can be considered to be at a uniform temperature, the direction of the radiation force is mostly upwards, and thus the wind flows initially vertically and can be considered a 1D wind. The corresponding equation describing the vertical motion is

\begin{equation}
    \rho \frac{\dd v_z}{\dd t} = \rho\, (a_\mathrm{grav}^z  + a_\mathrm{rad}^z) - \frac{\partial P}{\partial z},
\end{equation}
Even though the \textsc{Qwind} model does not include hydrodynamic forces when solving the 2D trajectories of gas parcels, we do include the force term due to gas pressure here, since it is necessary for deriving critical point like solutions (see Appendix \ref{app:initial_conditions}). 

The study of 1D line-driven winds was pioneered by \citetalias{castor_radiation-driven_1975}, who defined a framework to find steady state solutions of the 1D wind equation. Their methodology can be extended to any particular geometry of the gravitational and radiation fields, in particular, \cite{pereyra_steady_2006} (hereafter \citetalias{pereyra_steady_2006}) apply the \citetalias{castor_radiation-driven_1975} formalism to the study of cataclysmic variables (CVs). We here aim to further extend this approach to our case, by using the \citetalias{castor_radiation-driven_1975} formalism to calculate the properties of the 1D wind solutions from an accretion disk as initial conditions for the global 2D wind solution.

The core result of the \citetalias{castor_radiation-driven_1975} approach is that if a steady state solution of the 1D wind equation satisfies the following conditions:

\begin{itemize}
    \item the velocity increases monotonically with height,
    \item the wind starts subsonic,
    \item the wind extends towards arbitrarily large heights,
    \item the wind becomes supersonic at some height,
    \item the velocity gradient is a continuous function of position,
\end{itemize}
then the wind must pass through a special point called the critical point $z_c$, which can be derived without solving the wind differential equation, and thus the global properties of the wind such as its mass loss rate can be determined without resolving the full wind trajectory. To keep the main text concise, we refer the reader to Appendix \ref{app:initial_conditions} for a detailed derivation.

The previously specified conditions for the existence of a critical point solution may not be satisfied for all of the wind trajectories that we aim to simulate. For instance, a wind trajectory that starts in an upward direction and falls back to the disc because it failed to achieve the escape velocity does not have a velocity that increases monotonically with height. Furthermore, eventually the wind trajectory is no longer vertical, and the 1D approach breaks down. Having considered these possibilities, and only for the purpose of deriving the initial conditions of the wind, we assume that these conditions hold, so that we can derive the wind mass loss rate at the critical point, which we in turn use to determine the initial conditions of the wind. The full 2D solution of the wind may then not satisfy these conditions. 

\begin{figure}
    \centering
    \includegraphics[width=\columnwidth]{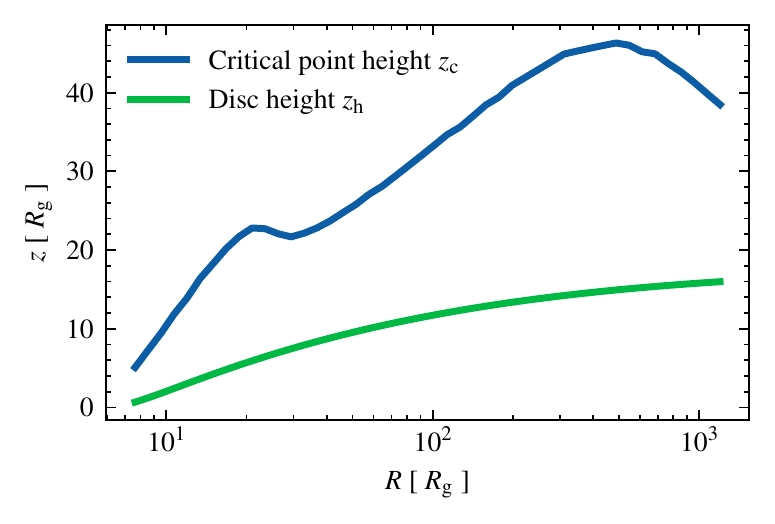}
    \caption{Position of the critical point $z_c$ as a function of radius, for $\M = 10^8 \msun$ and $\dot m =0.5$.}  
    \label{fig:critical_points}
\end{figure}

The location of the critical point as a function of radius is plotted in \autoref{fig:critical_points}, where we also plot the height of the disc,
\begin{equation}
    z_\mathrm{h}(R) = \frac{\sigmae \mu_e \mathcal F(R) R^3}{G\M c},
\end{equation}
defined as the point of equality between the vertical gravitational and radiation force. Overall, we notice that the critical point height increases slowly with radius, except for a bump at $R\sim 20\rg$ which is caused by the UV fraction dependence with radius (see \autoref{fig:uv_fractions}). For radii $R \lesssim 50\, \rg$ the critical point height is comparable to the disc radius, so our approximation that streamlines are vertical at that point may not be applicable. We assess the validity of this assumption in \autoref{subsection:verify_critical_point}. 

{}

\begin{figure}
    \centering
    \includegraphics[width=\columnwidth]{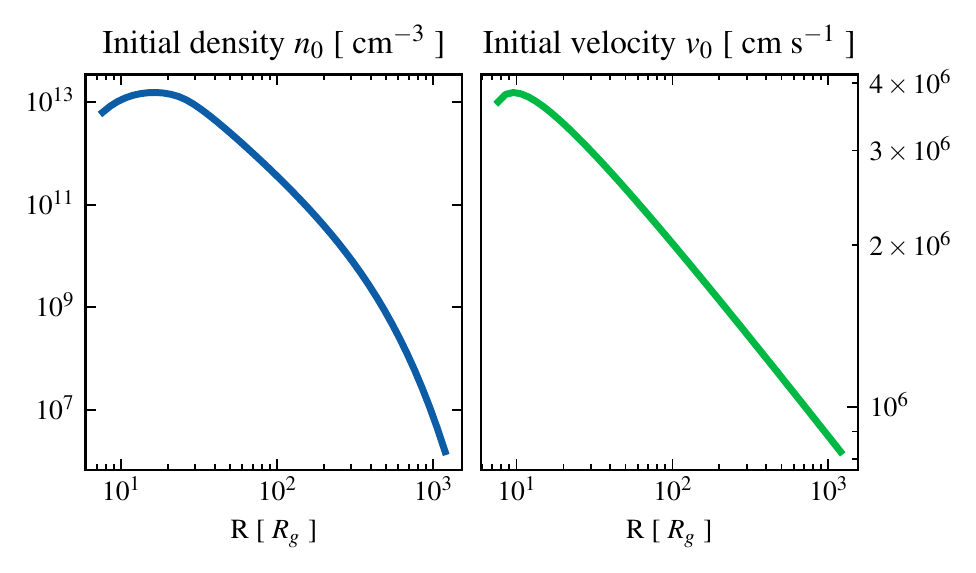}
    \caption{Initial conditions at the base of the wind as a function of radius. The original \textsc{Qwind} code assumed 
    constant initial density and velocity across the disk, and these were free parameters. We now calculate these from first principles. The left panel shows that the initial density changes by a factor of $10^6$, very different to the original assumption, while the right panel (note change in y axis scale) shows that the initial velocity changes by less than a factor 10. }  
    \label{fig:initial_conditions}
\end{figure}

We assume that the wind originates from the disc surface with an initial velocity $v_0$ equal to the thermal velocity (or isothermal sound speed) at the local disc temperature, 
\begin{equation}
    \label{eq:thermal_velocity}
    \vth(R) = \sqrt{\frac{k_\mathrm{B} T(R)}{\mu \, m_p}}.
\end{equation}
Given that the critical point is close to the disc surface, and that the wind is supersonic at the critical point (see Appendix \ref{app:initial_conditions}), this is a good starting point. Since mass conservation holds, the initial number density of the wind can then be calculated as
\begin{equation}
    n_0 (R) = \frac{\dot \Sigma (R)}{\vth(T(R)) \, \mu \, m_\mathrm{p}},
\end{equation}
where $\dot \Sigma(R)$ is the mass loss rate per unit area at the critical point. In \autoref{fig:initial_conditions}, we plot the initial number density and velocity for $\M =10^8\, \msun$, $\dot m =0.5$. We note that the initial velocity stays relatively constant, only varying by a factor of $\sim 3$ across the radius range. However, the initial number density varies by more than 5 orders of magnitude, showing that the assumption of a constant density at the base of the wind used in the previous versions of the model was poor. 

{}

\subsection{Comparison to other models}

\begin{figure}
    \centering
    \includegraphics[width=\columnwidth]{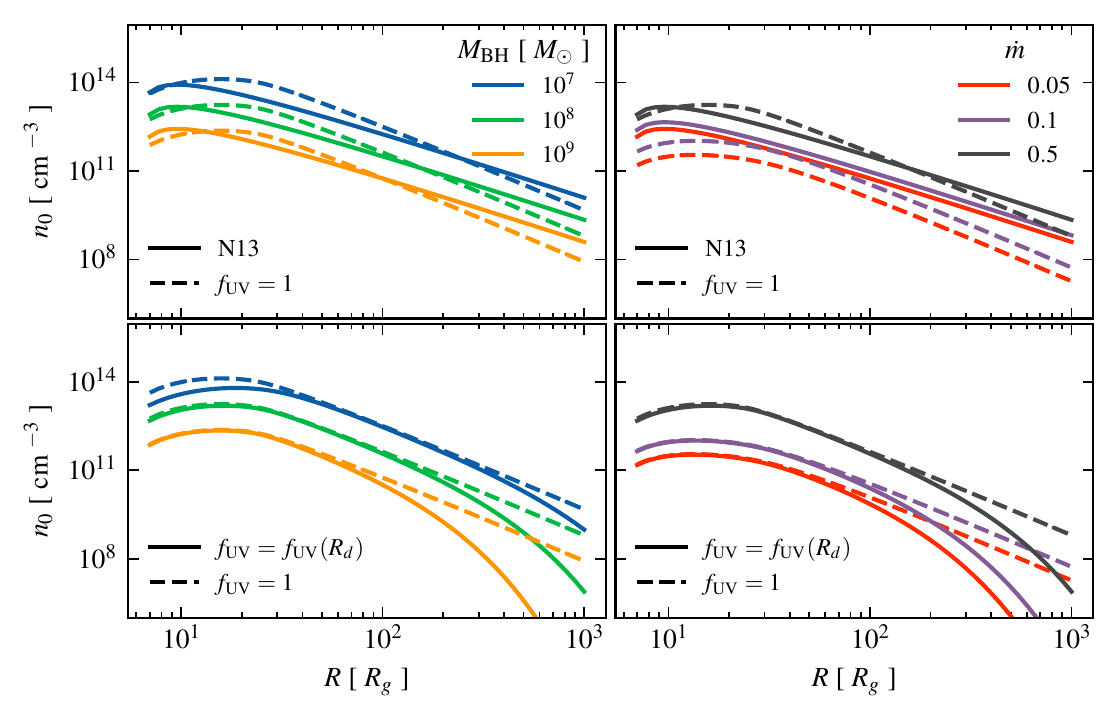}
    \caption{Initial density as a function of radius. In the top panels we compare with the results of \protect\citetalias{nomura_modeling_2013} (solid lines), the only other diskwind code which used UV line driving to calculate the initial density. That code assumes constant $f_{UV}$, so we fix $\fuv=1$ in  \textsc{Qwind} to compare results (dashed lines). In the bottom panels we compare the 
    \textsc{Qwind} results with $\fuv=1$ (dashed lines) with the full \textsc{Qwind} results for $\fuv(\rd)$. Left and right panels show  results for fixed $\dot m = 0.5$, and vary $\protect\M \in (10^7, 10^8, 10^9) \, \msun$. Right panels: We fix $\M = 10^8\,\msun$ and vary $\protect \dot m \in(0.05, 0.1, 0.5)$.}
    \label{fig:nomura_comparison}
\end{figure}

As a partial test of this new section of the code, we can compare our findings with \cite{nomura_modeling_2013} (hereafter \citetalias{nomura_modeling_2013}), in which the authors also use the sonic velocity as the initial velocity of the wind, and derive the initial density profile by assuming the same functional form for the mass loss rate as in \citetalias{castor_radiation-driven_1975}, but using the AGN $\M$ and $\dot m$ instead. We note that the use of the CAK formula directly for accretion discs may not be appropriate because the geometry of the system is very different from stellar winds. Additionally, in \citetalias{nomura_modeling_2013} the dependence of $\fuv$ on the disc radius is not considered. Hence we first hardwire $\fuv$ for the comparison. This comparison is shown in the upper panels of \autoref{fig:nomura_comparison} for different $\M$ (left, all at $\dot{m}=0.5$), and (right) for $\M$ fixed at $10^8\,\msun$ with different $\dot{m}$. It is clear that the initial density now derived in \textsc{Qwind3} (dashed lines) has a steeper decrease with radius than in \citetalias{nomura_modeling_2013}. This is probably due to their use of the direct CAK formula, which assumes a spherical geometry rather than an accretion disc. However, the inferred densities are within an order of magnitude of each other, and both approaches give a linear scaling of the initial number density profile with $\M$, but \textsc{Qwind3} gives an almost quadratic scaling with $\dot m$, compared to a linear one for \citetalias{nomura_modeling_2013} (see \autoref{subsection:ic_scaling}).

The lower panels of \autoref{fig:nomura_comparison} show instead the comparison of the \textsc{Qwind} models using the self consistent $\fuv$ with its radial dependence (solid lines) instead of assuming $\fuv=1$ (dashed lines). There is a very strong drop in the initial density at radii where $\fuv$ drops (see Fig. \ref{fig:uv_fractions}). This shows the importance of including the self-consistent calculation of $\fuv$ in \textsc{Qwind3}.

{}

\section{Updates to the radiation transport}
\label{sec:radiation_transport}

In \qo, the radiation transfer is treated in a very simple way. The disc atmosphere (i.e. the wind) is assumed to have constant density, and so the line of sight absorption does not take into consideration the full geometry and density structure of the wind (see section 2 of \qo). Furthermore, the UV optical depth is measured from the centre of the disc, and assumed to be the same for radiation from all disc patches, regardless of the position and angle relative to the gas parcel. In this section, we improve \qw's radiative transfer model, by reconstructing the wind density from the gas blob trajectories, thus accounting fully for the wind geometry. The disc is assumed to be flat and thin, with constant height $\bar z_\mathrm{h} = 0$, thus we do not model the effect of the disc itself on the radiation transfer. To illustrate the improvements, we consider our fiducial model with $\M=10^8\msun$, and $\dot m=0.5$, and present the ray tracing engine of the code in an arbitrary wind solution. We discuss particular physical implications and results in \autoref{sec:results}.

\begin{figure}

\tdplotsetmaincoords{75}{120}

\pgfmathsetmacro{\rvec}{.8}
\pgfmathsetmacro{\thetavec}{30}
\pgfmathsetmacro{\phivec}{60}

\begin{tikzpicture}[scale=3,tdplot_main_coords]

\coordinate (O) at (0,0,0);

\tdplotsetcoord{P}{\rvec}{\thetavec}{\phivec}

\draw[thick,->, dashed] (0,0,0) -- (1,0,0) node[anchor=north east]{$x'$};
\draw[thick,->, dashed] (0,0,0) -- (0,1,0) node[anchor=north west]{$y'$};
\draw[thick,->, dashed] (0,0,0) -- (0,0,1) node[anchor=south]{$z'$};

\fill[thick, dashed, fill=brown, opacity=0.1] (0,0,0) circle (1.0);

\tdplotsetcoord{disk_1}{0.7}{90}{120}
\tdplotsetcoord{disk_2}{0.7}{90}{130}
\tdplotsetcoord{disk_3}{0.8}{90}{120}
\tdplotsetcoord{disk_4}{0.8}{90}{130}
\tdplotsetcoord{disk_center}{0.75}{90}{125}
\tdplotsetcoord{disk_center_label}{0.45}{90}{155}
\tdplotsetcoord{disk_center_center}{0.75}{90}{125}
\draw (disk_1) -- (disk_2);
\draw (disk_2) -- (disk_4);
\draw (disk_1) -- (disk_3);
\draw (disk_3) -- (disk_4);
\draw (O) -- node[left] {} (disk_center) ;
\node (OO) at (disk_center_label) {$R_d$};
\tdplotdrawarc{(O)}{0.7}{0}{120}{anchor=north}{$\phi_d$}

\coordinate (gas) at (0.8,0,0.8);
\draw[thick, ->, violet] (disk_center_center) -- node[above right]{$\Delta$} (gas);
\draw[dashed] (0, 0, 0.8) -- node[above]{$R$} (gas);
\draw[dashed] (0.8, 0, 0.0) -- node[left]{$z$} (gas);
\draw[thick, ->, blue] (O) -- node[below left]{$r$} (gas);
\end{tikzpicture}
  \caption{Diagram of the disc-wind geometry. The blue line corresponds to the light path an X-ray photon takes, while the violet line corresponds to an example of a UV light ray from the accretion disc.}
  \label{fig:geometry_diagram}
\end{figure}
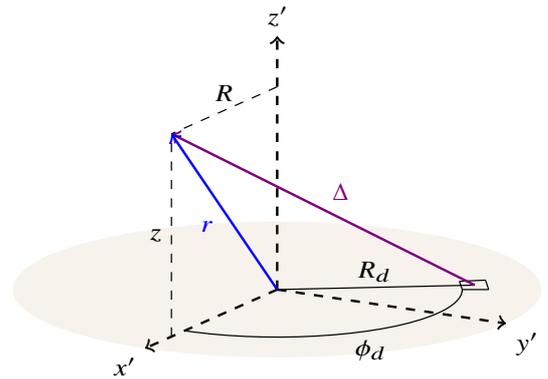

\subsection{Constructing the density interpolation grid}
\label{sec:interp_grid}

\begin{figure}
    \centering
    \includegraphics[width=\columnwidth]{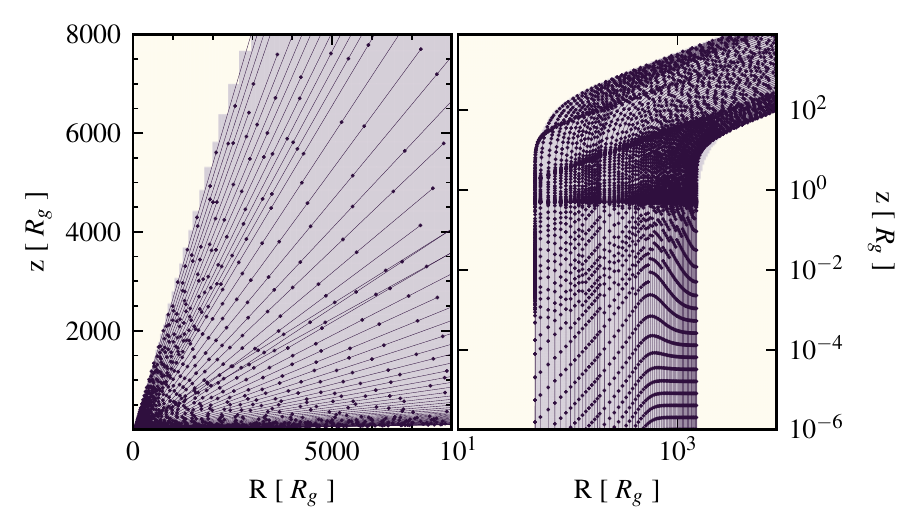}
    \caption{Gas parcel trajectories encapsulated by the concave hull containing the points. Left panel on linear scale and right panel on logarithmic scale. Note that the non-convexity of the wind prevents us from using a simpler convex hull envelope.}
    \label{fig:wind_hull}
\end{figure}

\begin{figure}
    \centering
    \includegraphics[width=\columnwidth]{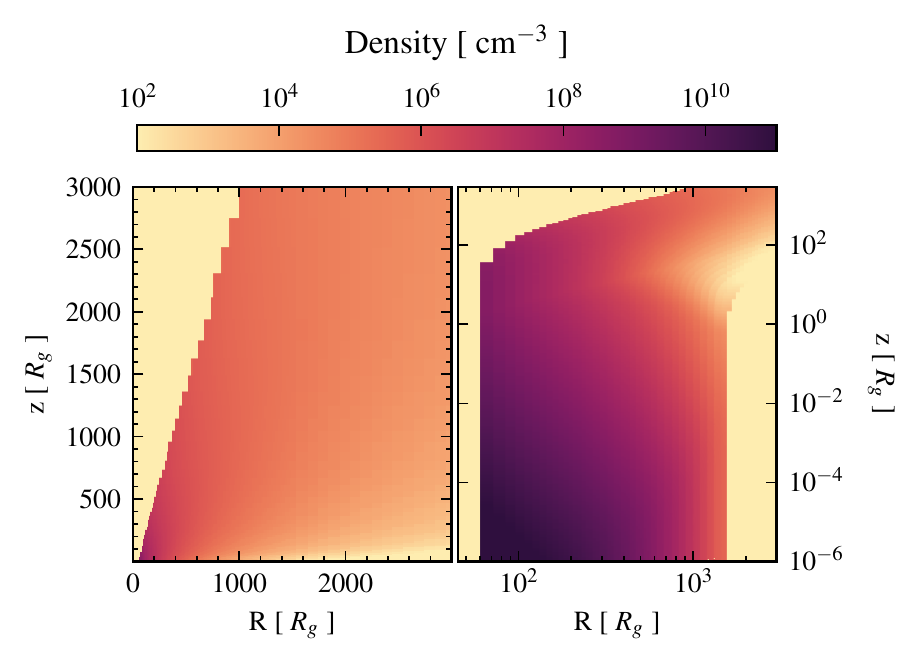}
    \caption{Interpolated density grid from the simulated wind positions and densities. Left panel: position in linear scale. Right panel: position in logarithmic scale.}
    \label{fig:density_grid}
\end{figure}

Given a collection of trajectories, we aim to obtain the wind density field at every point in space. The first step is to delimit where the wind is spatially located by computing the concave hull that contains all the points of all wind trajectories. We use the algorithm described in \cite{moreira_concave_2007} and implemented in \cite{stagner_lstagnerconcavehulljl_2021}. The resulting concave set is illustrated in \autoref{fig:wind_hull}. Outside the concave hull, the density is set to the vacuum density which is defined to be $n_\text{vac} = 10^2$ cm$^{-3}$. 
Since the density varies by orders of magnitude within the wind, we compute the density at a point by linearly interpolating $\log_{10} n$ in logarithmic space ($\log R$ - $\log z$) from the simulated wind trajectories. We use the interpolation algorithm \textsc{LinearNDInterpolator} from \textsc{SciPy} \citep{virtanen_scipy_2020}. The resulting density map is shown in \autoref{fig:density_grid}. We note that using a concave hull envelope is important, since the interpolation algorithm we use restricts the interpolation space to the convex hull of the input points, which, due to non-convexity of the wind geometry, would otherwise lead us to overestimate the obscuration in certain regions.

Once we have built the interpolator, we construct a rectilinear grid with the interpolated values. This allows us to implement an efficient ray tracing algorithm to compute the UV and X-ray optical depths. The vertical coordinates of the grid nodes are logarithmically spaced from $10^{-6}$ $R_g$ to the wind's maximum height. The horizontal coordinates are taken at the initial positions of the wind's trajectories, plus an additional range logarithmically spaced from the initial position of the last streamline to the highest simulated $R$ coordinate value.

\subsection{Ray tracing}

\begin{figure}
  \begin{tikzpicture}[scale=1]

\coordinate (O) at (0,0);
\coordinate (rectangle_max) at (5, 5);
\foreach \x in {0,...,4}{
    \draw[opacity=0.2, very thin] (\x,0) -- (\x, 5);
}
\foreach \y in {0,...,4}{
    \draw[opacity=0.2, very thin] (0,\y) -- (5, \y);
}

\foreach \x/\y/\n in {0/0, 0.888889/1.0, 1.0/1.125, 1.77778/2.0, 2.0/2.25, 2.66667/3.0, 3.0/3.375, 3.3333 /3.75}{
    \draw[fill=cyan] (\x, \y) circle[radius=2pt];
}
\draw[thick] (O) -- (3.3333, 3.75);
\draw[<->, thick, dashed] (1.15, 1.025) -- (1.9, 1.9);
\draw[<->, thick, dashed] (0.1, -0.2) -- (1.15, 1.025);

\node (A) at (1.85, 1.35) {$\Delta d_i$};
\node (B) at (1.0, 0.4) {$d_{0_i}$};
\node (C) at (3.3333, 3.75)[anchor=south west] {$(R, 0, z)$};
\node (n) at (0.90, 1.3)[anchor=north east] {$n_i$};


\definecolor{mycolor}{rgb}{0.588,0.09,0.768}

\foreach \x/\y/\n in {4.5/0.0, 4.0/0.309, 3.22/1.0, 3.0/1.5517241379310347, 3.0/1.7999999999999998, 3.0595932917809696/2.0, 4.0/3.0}{
    \draw[fill=mycolor] (\x, \y) circle[radius=2pt];
}

\draw[black, thick]   plot[smooth,domain=0:3] ({0.55556*\x^2 -1.8333*\x + 4.5}, \x);

\node (E) at (3.7, 2.4)[anchor=south west] {$(R', 0, z')$};

%
%
\node (OO) at (O)[anchor=north east] {$(0,0,0)$};
\node (OOO) at (4.5, 0)[anchor=north] {$(R_d,\phi_d, 0)$};
\draw (O) rectangle (rectangle_max);

\end{tikzpicture}
  \caption{Projections of two typical light rays onto the $R-z$ interpolation grid. The X-ray radiation (path with blue dots) is assumed to come from the centre of the grid $(0,0)$, so the projection of the light curve onto the $R-z$ grid is always a straight line. However, for the UV case (purple dots), the light ray can originate from any $\pd$, so the path on the interpolation grid is, in general, a parabola.}
  \label{fig:tikz:ray_tracing}
\end{figure}
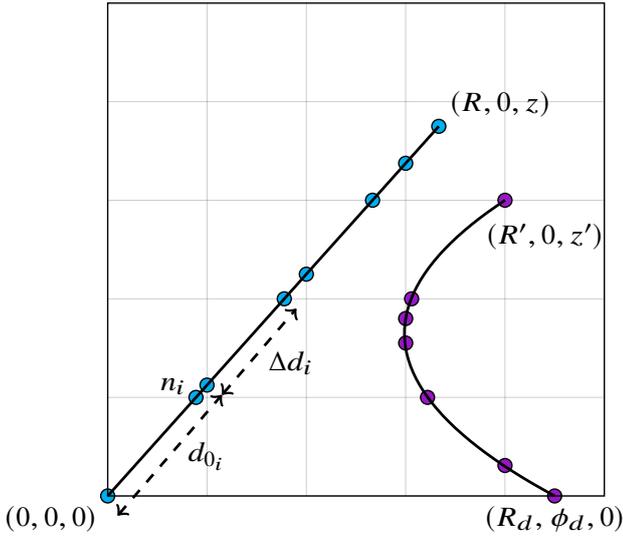

To compute the optical depth along different lines of sight, we need to calculate an integral along a straight path starting at a disc point $(\rd, \pd, 0)$ to a point $(R, \phi, z)$. Due to the axisymmetry of the system, the radiation acceleration is independent of $\phi$ so we can set $\phi = 0$. We can parametrise the curve in the Cartesian coordinate system with a single parameter $t\in [0,1]$, (see \autoref{fig:geometry_diagram}),
\begin{equation}
    \begin{split}
        x(t) &= \rd \cos\pd \, (1-t) + t \, R , \\
        y(t) &= \rd \sin\pd \, (1-t),\\
        z(t) &= t \, z\\
    \end{split}
\end{equation}
so that the cylindrical radius varies along the path as
\begin{equation}
    R_t^2(t) = \rd^2(1-t)^2 + t^2 R^2 + 2\rd R \cos\pd \,t (1-t) ,
\end{equation}
and $\phi(t) = \arctan{(y(t) / x(t))}$. In this parametrisation, $t=0$ points to the disc plane point, and $t=1$ to the illuminated wind element. The integral to compute is thus
\begin{equation}
    \tau = \Delta \, \int_0^1 \sigma(t)\; n(t)\, \dd t,
\end{equation}
where $\Delta$ is the total path length defined earlier.
Given a rectilinear density grid in the R-z plane, we compute the intersections of the light ray with the grid lines, $\{ R_i, z_i\}$, such that we can discretise the integral as,
\begin{equation}
    \tau \approx \sum_i \sigma(R_i, z_i) \, n(R_i, z_i)\Delta d_i,
\end{equation}
where $\Delta d_i$ is the 3D distance between the $i$-th intersection point and the $(i-1)$-th,
\begin{equation}
    \Delta d_i = \sqrt{R_i^2 + R_{i-1}^2 + (z_i - z_{i-1})^2 -2 R_i R_{i-1} \cos(\phi_i-\phi_{i-1})},
\end{equation}
To find the intersections, we need to calculate whether the light ray crosses an $R_i$ grid line or a $z_i$ grid line. We start at the initial point $(\rd, \pd, 0)$, and compute the path parameter $t_R$ to hit the next $R$ grid line $R_i$ by solving the second degree equation $R(t_R) = R_i$, and similarly for the next $z_i$ line, $t_z = z_i / z$. The next intersection is thus given by the values of $R(t_m)$ and $z(t_m)$ where $t_m = \min(t_R, t_z)$. Geometrically, the projection of the straight path onto the $(R-z)$ grid is in general a parabola as we can see in \autoref{fig:tikz:ray_tracing}. 

\subsection{X-ray optical depth}

\begin{figure}
    \centering
    \includegraphics[width=\columnwidth]{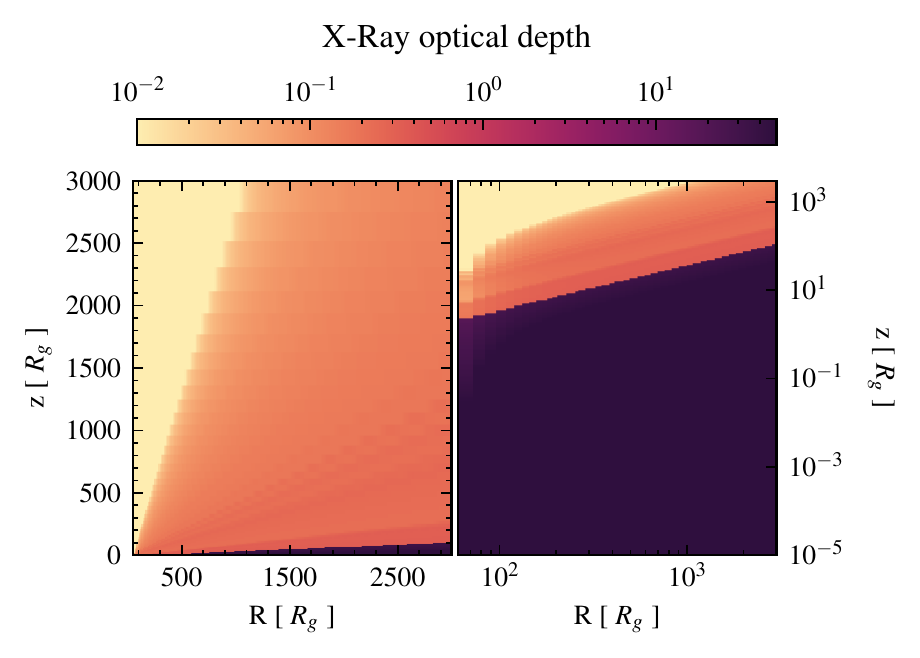}
    \caption{X-ray optical depth as a function of position, measured from $R=0$ and $z=0$. Left panel: position in linear scale. Right panel: position in logarithmic scale.}
    \label{fig:xray_grid}
\end{figure}

The X-ray opacity depends on the ionisation level of the gas and is assumed to have the same functional form as \qo,
\begin{equation}
    \label{eq:xray_opacity}
    \sigmax(\xi) = \begin{cases}\sigmat & \text{ if } \xi > 10^5 \text{erg cm s}^{-1}\\ 100 \sigmat & \text{ if }\xi \leq 10^5 \text{erg cm s}^{-1}\end{cases},
\end{equation}
where $\sigmat$ is the Thomson scattering cross section, $\xi$ is the ionisation parameter,
\begin{equation}
    \xi = \frac{4\pi F_\text{\tiny X}}{n},
\end{equation}
and $F_\text{\tiny X}$ is the X-ray radiation flux, $F_\text{\tiny X} = L_\text{\tiny X} \exp(-\taux) / (4\pi r^2)$. We notice that to compute the value of $\taux$ we need to solve an implicit equation, since the optical depth depends on the ionisation state of the gas which in turn depends on the optical depth. One thus needs to compute the distance $d_\text{\tiny X}$ at which the ionisation parameter drops below $\xi_0 = 10^5 \text{ erg cm s}^{-1}$,
\begin{equation}
    \xi_0 - \frac{L_\text{\tiny X}}{n d_\text{\tiny X}^2} \exp{(-\taux)} = 0.
\end{equation}
This equation needs to be tested for each grid cell along the line of sight as it depends on the local density value $n$. Therefore for a cell at a distance $d_{0_i}$ from the centre, density $n_i$, and intersection length $\Delta d_i$ with the light ray (see \autoref{fig:tikz:ray_tracing}), the contribution $\Delta \taux$ to the optical depth is
\begin{equation}
    \Delta \taux = \sigmat \, n_i \cdot \left[\max(0, d_\text{\tiny X}-d_{0_i}) + 100 \cdot \max(0, \Delta d_i - (d_\text{\tiny X} - d_{0_i}))\right],
\end{equation}
where $d_\text{\tiny X}$ is calculated from
\begin{equation}
    \xi_0 - \frac{L_\text{\tiny X}}{n_i d_\text{\tiny X}^2} \exp{\left(-\tau_{0_i} - \sigmat \cdot n_i \cdot (d_\text{\tiny X} - d_{0_i})\right)} = 0,
\end{equation}
where $\tau_{0_i}$ is the accumulated optical depth from the centre to the current position. We solve the equation numerically using the bisection method. In \autoref{fig:xray_grid} we plot the X-ray optical depth grid for our example wind. We observe that there is a region where $\taux \gg 1$ at very low heights. This shadow is caused by the shielding from the inner wind, which makes the ionisation parameter drop below $\xi_0$, substantially increasing the X-ray opacity. As we will later discuss in the results section, the shadow defines the acceleration region of the wind, where the force multiplier is very high.

\subsection{UV optical depth}

\begin{figure}
    \centering
    \includegraphics[width=\columnwidth]{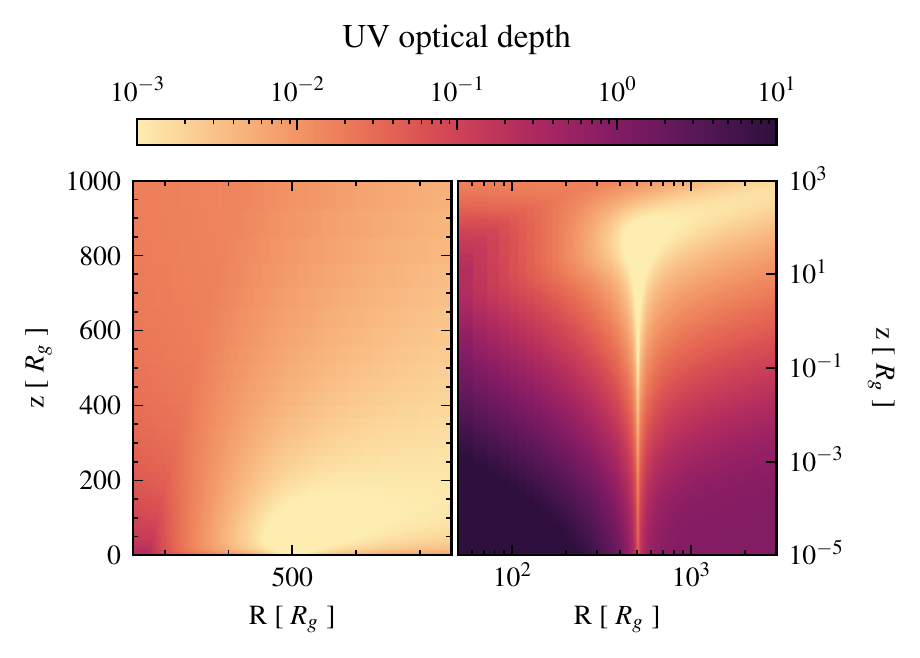}
    \caption{UV optical depth as a function of position, measured from $R=500\rg$ and $z=0$. Left panel: position in linear scale. Right panel: position in logarithmic scale.}
    \label{fig:uv_grid}
\end{figure}

The UV opacity calculation is significantly simpler than that for the X-rays, since we  assume that the line shift due to the Doppler effect in an accelerating wind is sufficient to always reveal fresh, unabsorbed continuum, so that the opacity is constant at the Thomson (electron scattering) value, $\sigma(R_i, z_i) = \sigmat$. 
In \autoref{fig:uv_grid}, we plot the UV optical depth as a function of $R$ and $z$ for light rays originating at the disc position $\rd = 500$, $\pd=0$. Nevertheless, there are many more sight-lines to consider as the UV emission is distributed over the disc, making this ray tracing calculation the highest contributor to the computational cost of the model. The total UV flux and its resultant direction at any given position in the wind have to be calculated as the sum over each disc element (see \autoref{eq:radiation_flux}) where now $\tauuv = \tauuv(R_d, \phi_d, R, z)$.

As already noted in \qo, the integral in \autoref{eq:radiation_acceleration} is challenging to calculate numerically, and in this case the computational cost is further increased by the refined UV ray tracing. In spite of that, by using the adaptive integration scheme presented in \cite{berntsen_adaptive_1991} and implemented in \cite{johnson_juliamathcubaturejl_2021}, the integration typically converges after $\mathcal O(10^4)$ integrand evaluations, which results in a computation time of a few milliseconds, keeping the simulation tractable. At low heights, $z\sim 0$, the trajectory solver requires several evaluations of the radiation force to correctly compute the adaptive time-step, which makes the computation particularly slow. Fortunately, the approximation in \autoref{eq:radiation_acceleration_approx} comes in very handy at reducing the overall computational cost at low heights.

\section{Special relativity effects}
\label{sec:relativistic}
When the gas trajectory approaches the speed of light, one should consider special relativistic effects such as relativistic beaming and Doppler shifting \citep[see eg][chapter 4]{rybicki_radiative_1986}. The importance of taking these effects into account is highlighted in \cite{luminari_importance_2020}. We include a correction to the radiation flux seen by the gas (\autoref{eq:radiation_flux}),
\begin{equation}
    \dd F_\text{relativistic}= \Psi(R_d, \phi_d, R, z, v_R, v_z) \; \dd F,
\end{equation}
where $v_R$ and $v_z$ are the radial and vertical velocity components of the gas at the position $(R, 0, z)$. We ignore the contribution from the angular velocity component, $v_{\phi}$ for simplicity, as its inclusion would break our assumption that angular momentum is conserved along a gas blob trajectory. The correction $\Psi$ is given by \citep{luminari_speed_2021},
\begin{equation}
    \Psi = \frac{1}{\gamma^4 (1+\beta\cos\theta)^4},
\end{equation}
where $\gamma$ is the Lorentz factor, $\beta = \sqrt{v_R^2 + v_z^2} / c$, and $\theta$ is the angle between the incoming light ray and the gas trajectory,
\begin{equation}
    \cos\theta = \frac{(R - R_d\cos\phi_d) v_R + z v_z}{\beta \Delta}.
\end{equation}
Intuitively, when the incoming light ray is parallel to the gas trajectory, $\cos\theta=1$, so the correction reduces to $\Psi = \left(\frac{1-\beta}{1+\beta}\right)^2$, which is 0 when $\beta=1$ and 1 when $\beta=0$, as expected. 

It is worth noting that this is a local correction which needs to be integrated along all the UV sight-lines (see \autoref{eq:radiation_acceleration}). Nonetheless, the computational cost of calculating the radiation force is heavily dominated by ray tracing and the corresponding UV optical depth calculation, so this relativistic correction does not significantly increase the computation time. 

The X-ray flux, which determines the ionisation state of the gas, is also likewise corrected for these special relativistic effects, but there is only one such sight-line to integrate along for any position in the wind, as the X-ray source is assumed to be point-like.

\section{Calculating the gas trajectories}
\label{sec:gas_trajectories}
\subsection{Initial radii of gas trajectories}

The first thing to consider when calculating the gas trajectories is the initial location of the gas blobs. The innermost initial position of the trajectories is taken as a free parameter $R_\mathrm{in}$, and the outermost initial position, $R_\mathrm{out}$, is assumed to be the self-gravity radius of the disc, where the disc is expected to end \citep{laor_massive_1989}, which for our reference BH corresponds to 1580 $R_g$. We initialise the first trajectory at $\rin$, the next trajectory starts at $R=R_\mathrm{in} + \Delta R$, where $\Delta R$ is the distance between adjacent trajectories. We determine $\Delta R$ by considering two quantities: (i) the change in optical depth between two adjacent trajectories along the base of the wind and (ii) the mass loss rate along a trajectory starting at $R$ and at a distance $\Delta R$ to the next one. Regarding (i), the change in optical depth $\Delta\tau$ between two trajectories initially separated by $\Delta R$ is given by
\begin{equation}
    \Delta\tau = \int_R^{R+\Delta R_1} n(R')\, \sigmat \,\dd R'.
\end{equation}
We consider
\begin{equation}
    \Delta\tau = \begin{cases} 
    0.05 & \mathrm{ if } \;\tau(R) < 5,\\
    0.5 & \mathrm{ if } \;\tau(R) < 10,\\
    5 & \mathrm{ if } \;\tau(R) < 100,\\
    20 & \mathrm{ if } \;\tau(R) > 100,\\
    \end{cases}
\end{equation}
where $\tau(R) = \sum_i \Delta\tau_i$. This guarantees that the spacing between trajectories resolves the transition from optically thin to optically thick for both the UV and X-ray radiation. The quantity (ii) is given by
\begin{equation}
    \Delta \dot M_\mathrm{wind} = \int_R^{R+\Delta R_2}2\pi R'\rho(R')v(R') \dd R',
\end{equation}
where we consider $\Delta \dot M_\mathrm{wind}=0.01\dot M$, so that no streamline represents more than $1\%$ of the accreted mass rate. The $\Delta R$ step is thus given by 
\begin{equation}
    \Delta R = \min (\Delta R_1, \Delta R_2)
\end{equation}
and we repeat this process until $R=R_\mathrm{out}$

\subsection{Solving the equation of motion}

The equation of motion of the wind trajectories is the same as in \qo, and we solve it analogously by using the Sundials IDA integrator \citep{hindmarsh_sundials_2005}. The wind trajectories are calculated until they either exceed a distance from the centre of $10^4 R_g$, fall back to the disc, or self intersect. To detect when a trajectory self intersects we use the algorithm detailed in Appendix \ref{app:intersections}.

\begin{figure}
    \centering
    \includegraphics{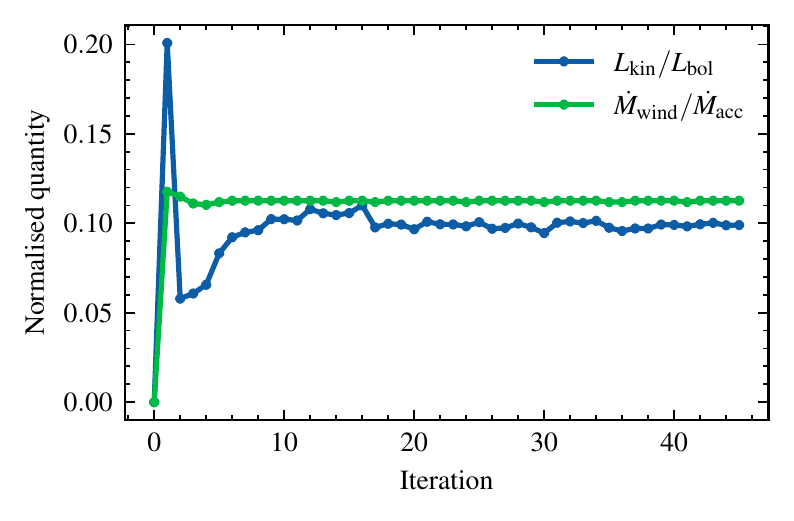}
    \caption{Normalised mass loss rate and kinetic luminosity for each iteration for our fiducial case.}
    \label{fig:iterations}
\end{figure}

By considering the full wind structure for the radiation ray tracing, we run into an added difficulty: the interdependence of the equation of motion with the density field of the wind. To circumvent this, we adopt an iterative procedure in which the density field of the previous iteration is used to compute the optical depth factors for the current iteration. We first start assuming that the disc's atmosphere is void, with a vacuum density of $n_\text{vac} = 10^2$ cm$^{-3}$. Under no shielding of the X-ray radiation, all the trajectories fall back to the disc in a parabolic motion. After this first iteration, we calculate the density field of the resulting failed wind and use it to calculate the wind trajectories again, only that this time the disc atmosphere is not void. We keep iterating until the mass loss rate and kinetic luminosity do not significantly change between iterations. It is convenient to average the density field in logarithmic space between iterations, that is, the density field considered for iteration $k$, $n_k$, is given by
\begin{equation}
    \logt(n_{k}(R, z)) = \frac{1}{2}\, \left(\logt(n_{k-1}(R,z)) + \logt(n_{k-2}(R,z))\right).
\end{equation}
We do not take the average for the first two iterations. The number of iterations required depends on the initial radius of the wind, but we typically find convergence after $\sim 20$ iterations, although we run several more to ensure that the standard deviation of the density field between iterations is small. The normalised mass loss rate and kinetic luminosity for our fiducial model at each iteration are shown in \autoref{fig:iterations}.

For a given iteration, solving the equation of motion for the different gas trajectories is an embarrassingly parallel problem, and so our code's performance scales very well upon using multiple CPUs, allowing us to quickly run multiple iterations, and scan the relevant parameter spaces. The computational cost of running one iteration is $\sim 5$ CPU hours, so one is able to obtain a fully defined wind simulation after $\sim 100$ CPU hours.

\section{Effects of the code improvements on the wind}
In this section, we evaluate the effect on the wind of the three improvements that we presented to the treatment of the radiation field: (i) radial dependence of $\fuv$, (ii) improved radiation transport, and (iii) relativistic corrections.
\begin{figure*}
    \centering
    \includegraphics[width=\textwidth]{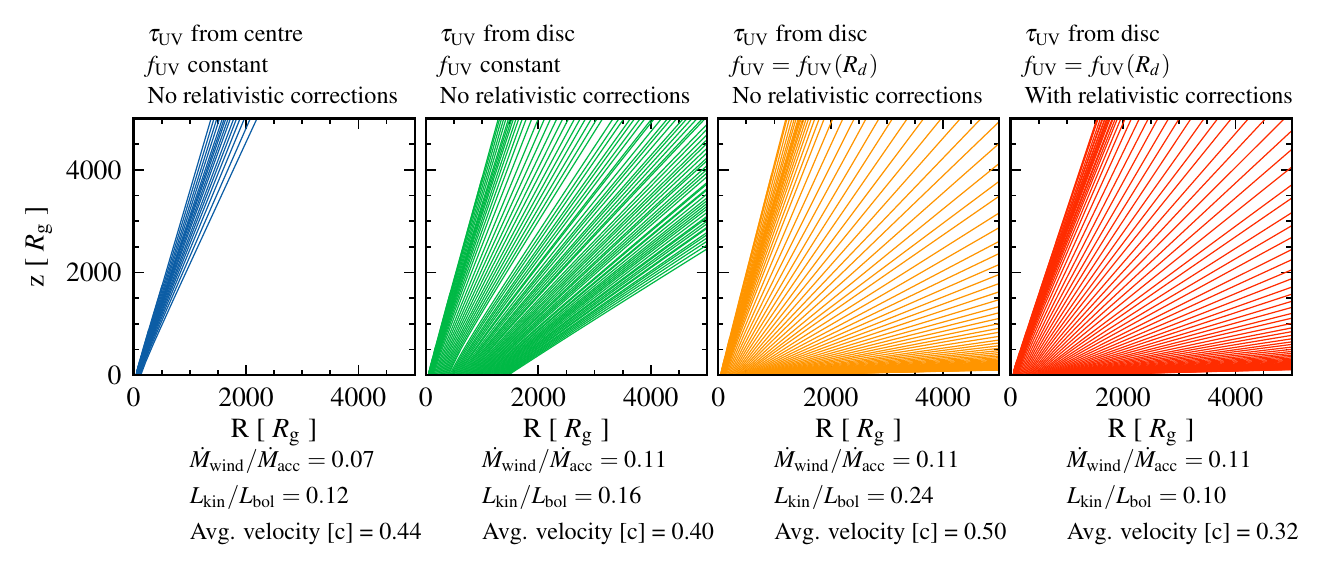}
    \caption{Effects of our improvements to the treatment of the radiation field on the trajectories of gas blobs. All simulations have been done with $\M = 10^8\msun$, $\dot m =0.5$, and $\rin=50\rg$.}
    \label{fig:comparisons}
\end{figure*}
The impact of each improvement is shown in \autoref{fig:comparisons}. In the first panel (blue), we calculate the optical depth $\tauuv$ from the centre, rather than attenuating each individual light ray coming from the disc. We also take a constant $\fuv=0.85$ and we do not include relativistic corrections. Calculating $\tauuv$ from the centre overestimates the attenuation of the UV radiation field in the outer parts of the wind, since it assumes that the radiation is crossing all the inner gas. This produces a failed wind in the outer regions of the disc (which is on too small a scale to be seen in \autoref{fig:comparisons}), since the inner gas is optically thick to the UV radiation. In the next panel (green), we calculate $\tauuv$ by attenuating each individual light ray as explained in \autoref{sec:radiation_transport}, consequently, most of the wind escapes from the disc since the UV attenuation is less strong, with the average wind velocity being slightly lower because we are now averaging over the slower trajectories in the outer part of the wind. The next improvement we evaluate is the inclusion of the radial dependence of $\fuv$ (third panel, orange). The change in wind geometry can be explained by considering the distribution of the UV emissivity, which is skewed towards the centre of the disc thus pushing the outer wind along the equator. This also leads to a more powerful wind, due to the increase in UV luminosity at small radii, where the fastest and most massive streamlines originate. Lastly, the effect of including relativistic corrections is shown in the 4th panel (red). As expected, the velocity of the wind is considerably lower, and hence also its kinetic luminosity. Furthermore, the wind flows at a slightly higher polar angle, as the vertical component of the radiation force gets weaker where the wind has a high vertical velocity. 
\section{Results}
\label{sec:results}

Here, we evaluate the dependence of the wind properties on the initial wind radius, BH mass, and mass accretion rate. We also study the impact of the relativistic corrections on the wind velocity and structure. All of the parameters that are not varied are specified in \autoref{table:fixed_parameters}.

\begin{table}
\centering
\begin{tabular}{ c c c }
 Parameter & Value \\ 
 \hline\hline 
 $f_\mathrm{\tiny X}$ & 0.15 \\
 $z_0$ & 0 \\
 $R_\text{out}/ \rg$ & 1580 \\
 $\mu$ & 0.61 \\  
 $\mu_e$ & 1.17 \\  
 $\alpha$ & 0.6  \\
 $k_\text{ic}$ & 0.03 \\
\end{tabular}
\caption{\label{table:fixed_parameters} Fixed parameters for the results section. Note that $k_\text{ic}$ refers to the value of $k$ used to compute the initial conditions of the wind (\autoref{eq:fm_simple}), but we use the SK90 parametrisation $k=k(\xi)$ elsewhere.}
\end{table}

\subsection{The fiducial case}

To gain some intuition about the structure of the wind trajectories solutions, we first have a close look at our fiducial simulation with $\M = 10^8\, \msun$, $\dot m =0.5$, and $\rin = 50\,\rg$. We run the simulation iterating 50 times through the density field, to make sure that our density grid has converged (see \autoref{sec:interp_grid}). 

\begin{figure}
    \centering
    \includegraphics[width=\columnwidth]{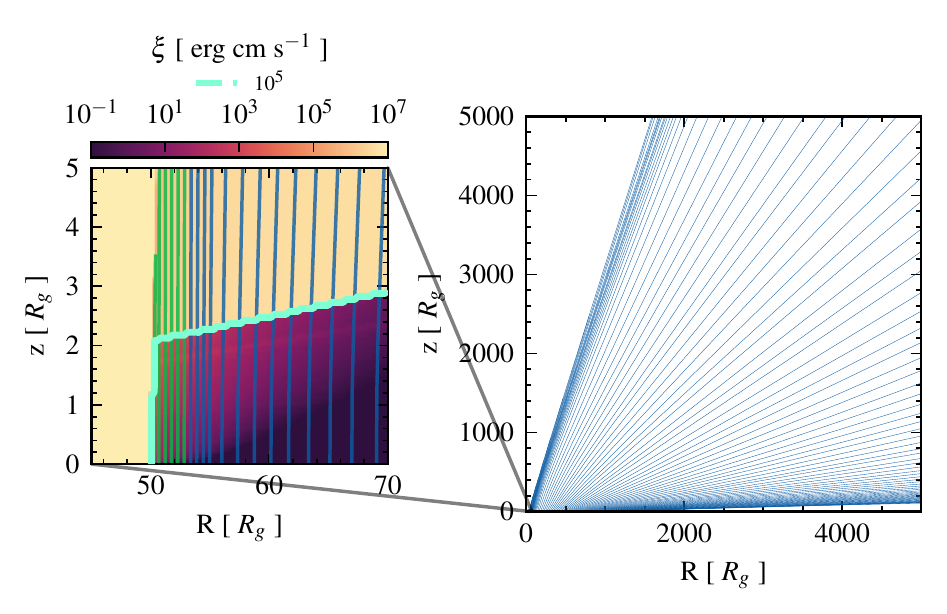}
    \caption{Simulated wind trajectories for our fiducial system ($\M = 10^8\msun$, $\dot m =0.5$, and $\rin = 50 \rg$) zooming in on the failed wind region, where we also colour-plot the ionisation parameter $\xi$, showing the contour at $\xi = 10^5$ erg cm s$^{-1}$ as the light turquoise line. We plot the failed trajectories in green and the escaping trajectories in blue.}
    \label{fig:failed_wind}
\end{figure}

In \autoref{fig:failed_wind}, we plot the wind streamline shapes, zooming in on the innermost region where we also show the ionisation state of the gas. \autoref{fig:initial_conditions} shows that the initial density should be $\sim 3\times 10^{12}$~cm$^{-3}$. Hence the initial ionisation parameter at $\rin$ is $\xi=f_\text{\tiny X}\, 0.5\, L_\mathrm{Edd}/(n_\mathrm{in} \rin^2)\sim 800$ so the base of the wind is already in the regime where the X-ray opacity is high. 
The wind starts, but the drop in density as the material accelerates means it reaches a high ionisation parameter where the force multiplier is low before it reaches escape velocity. Hence the material falls back to the disc as a failed wind region. 

The failed wind region has a size characterised by $\tau_x \lesssim 5$, acting as a shield to the outer wind from the central X-ray source. The X-ray obscuration is especially large in the failed wind shadow, due to the jump in X-ray opacity at $\xi_0 = 10^5$ erg cm s$^{-1}$ (see contour shown by turquoise line in the left panel of \autoref{fig:failed_wind}), but its opening angle may be small (\autoref{fig:xray_grid}). The shadow region defines the acceleration region of the wind, where the force multiplier is greatly enhanced and the wind gets almost all of its acceleration. Eventually this acceleration is enough that the material reaches the escape velocity before it emerges from the shadow, and is overionised by the X-ray radiation. The left panel of 
Fig. \ref{fig:failed_wind} shows these first escaping streamlines (blue) which are close to $\rin$. 

\begin{figure}
    \centering
    \includegraphics[width=\columnwidth]{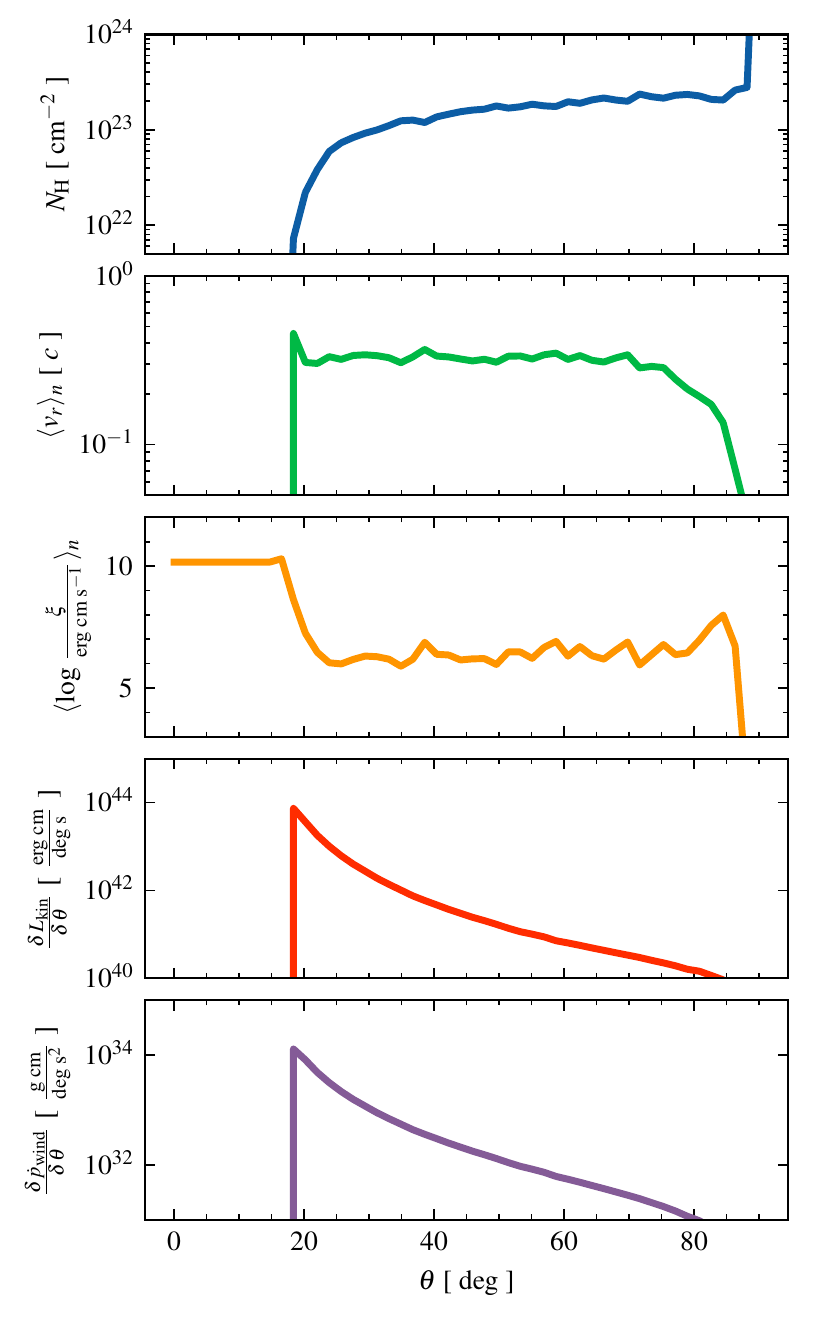}
    \caption{Wind properties for a system with $\M = 10^8\msun$, $\dot m =0.5$, $\rin=50\rg$ measured along a sight-line at angle $\theta$ at a distance $r=5000\rg$ from the centre. First panel: column density. Second panel: outward mean velocity weighted by density. Third panel: mean ionisation parameter weighted by density. Fourth panel: kinetic luminosity per unit angle. Fifth panel: wind momentum rate per unit angle.}
    \label{fig:ufo_pros}
\end{figure}

\autoref{fig:ufo_pros} shows the resulting wind parameters at 
a distance $r=5000 \, \rg$. We plot the wind column density, and density-weighted mean velocity and ionisation parameter as a function of the polar angle $\theta = \arctan(R/z)$. The column is almost constant at $N_H\sim 2\times 10^{23}$~cm$^{-2}$ (optical depth of $\sim 0.1$ to electron scattering) across the range $25^\circ<\theta<85^\circ$. For $\theta > 85^\circ$, the sight-line intercepts the inner failed wind and the column density increases to $N_H\sim 10^{24}$~cm$^{-2}$. The
typical wind velocity at this point is  $\simeq (0.1-0.4)\,c$
but it is always very ionised ($\xi > 10^5$ erg cm s $^{-1}$) at these large distances. This is too ionised to allow even H- and He-like iron to give visible atomic features in this high velocity gas, although these species may exist at smaller radii where the material is denser. We will explore the observational impact of this in a future work, specifically assessing whether UV line-driving can be the origin of the ultra-fast outflows seen in some AGN
(see also \citealt{mizumoto_uv_2021}). 

The escaping wind carries a mass loss rate of $\dot M_\mathrm{wind} \simeq 0.26$ $\msun /$ yr, corresponding to $\dot M_\mathrm{wind} / \dot M \simeq 11\%$ of the mass accretion rate, and a kinetic luminosity of $L_\mathrm{kin} \approx 7 \times 10^{44}$ erg / s, which is equal to 10\% of the bolometric luminosity. As the two bottom panels of \autoref{fig:ufo_pros} show, most of the energy and momentum of the wind is located at small polar angles ($\sim 20^\circ$), which is consistent with the initial density profile since the innermost streamlines carry the largest amount of mass.

\subsection{Dependence on the initial radius \texorpdfstring{$\rin$}{Rin}}

\begin{figure*}
    \centering
    \includegraphics[width=\textwidth]{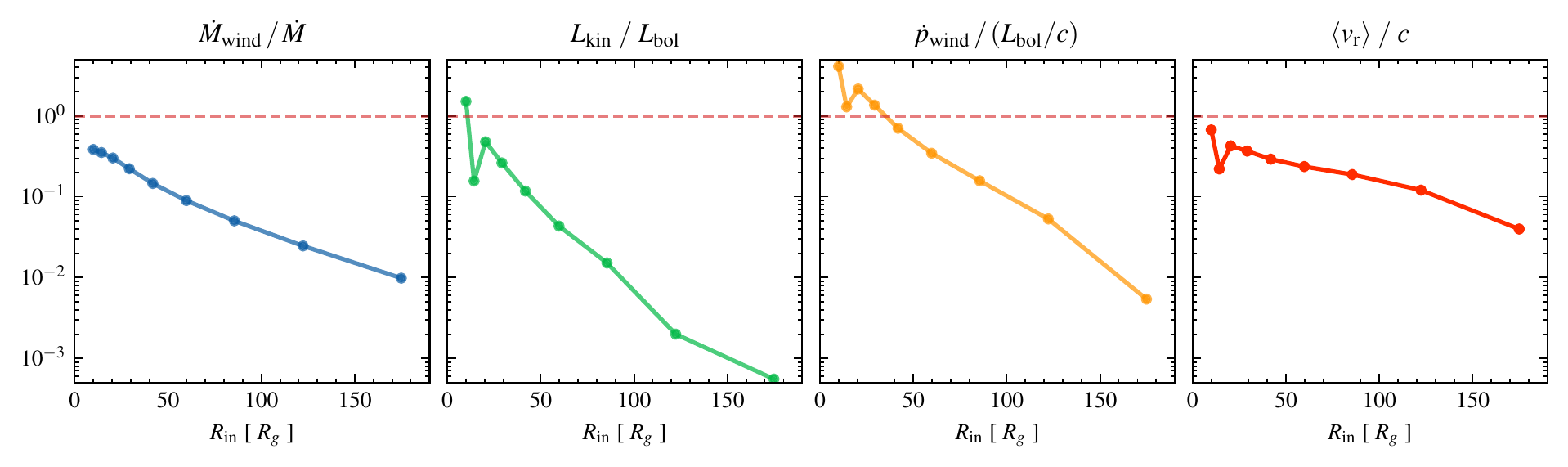}
    \caption{Mass loss rate (first panel), kinetic luminosity (second panel), momentum loss rate (third panel) and average velocity (fourth panel) for different values of $\rin$, at a fixed $\M=10^8\msun$ and $\dot m =0.5$. The mass loss rate is normalised to the system's mass accretion rate, while we normalise the luminosity to the bolometric luminosity. The average velocity is taken as \protect$v_\mathrm{r} = \sqrt{2 L_\mathrm{kin} / \dot M_\mathrm{wind}}$.}
    \label{fig:rin_scan}
\end{figure*}

As we already mentioned in \autoref{sec:gas_trajectories}, the initial radius of the innermost trajectory ($\rin$) is left as a free parameter to explore. This parameter is likely dependent on the structure of the accretion flow, which we do not aim to model here. As we increase $\rin$, the amount of mass that can potentially be lifted from the disc decreases, both because of the reduction in the extent of the launching region and the decrease in initial density with radius (\autoref{fig:initial_conditions}). Furthermore, increasing $\rin$ also narrows the failed wind shadow, since it reduces its subtended angle, thus reducing the accelerating region of the wind. We therefore expect the wind to flow at higher polar angles and smaller velocities when increasing $\rin$. In \autoref{fig:rin_scan}, we plot the predicted normalised mass loss rate, kinetic luminosity, momentum loss rate, and average velocity of the wind as a function of $\rin$. As we expected, both the mass loss rate and the kinetic luminosity decrease with $R_\mathrm{in}$, with the latter decreasing much faster. This difference in scaling is not surprising, since the fastest part of the wind originates from the innermost part of the disc, where most of the UV radiation is emitted. To further illustrate this, we plot the average velocity of the wind for each simulation in the rightmost panel of \autoref{fig:rin_scan}, observing that the maximum velocity decreases with initial radius. We note that the wind successfully escapes for $\rin \gtrsim 175 \rg$, which is a consequence of the initial number density profile (\autoref{fig:initial_conditions}) sharply declining after $R \gtrsim 100 \rg$. There is a physically interesting situation happening at $\rin \sim 12\rg$, where the average velocity of the wind drops. This is due to the failed wind being located where most of the UV emission is, thus making the wind optically thick to UV radiation (with respect to electron scattering opacity). Lastly, we note that the wind consistently reaches velocities $\gtrapprox 0.3$ $c$ for $\rin \lesssim 50\, \rg$.

\subsection{Dependence on BH mass and mass accretion rate}
\label{sec:results_bh}

\begin{figure*}
    \centering
    \includegraphics[width=\textwidth]{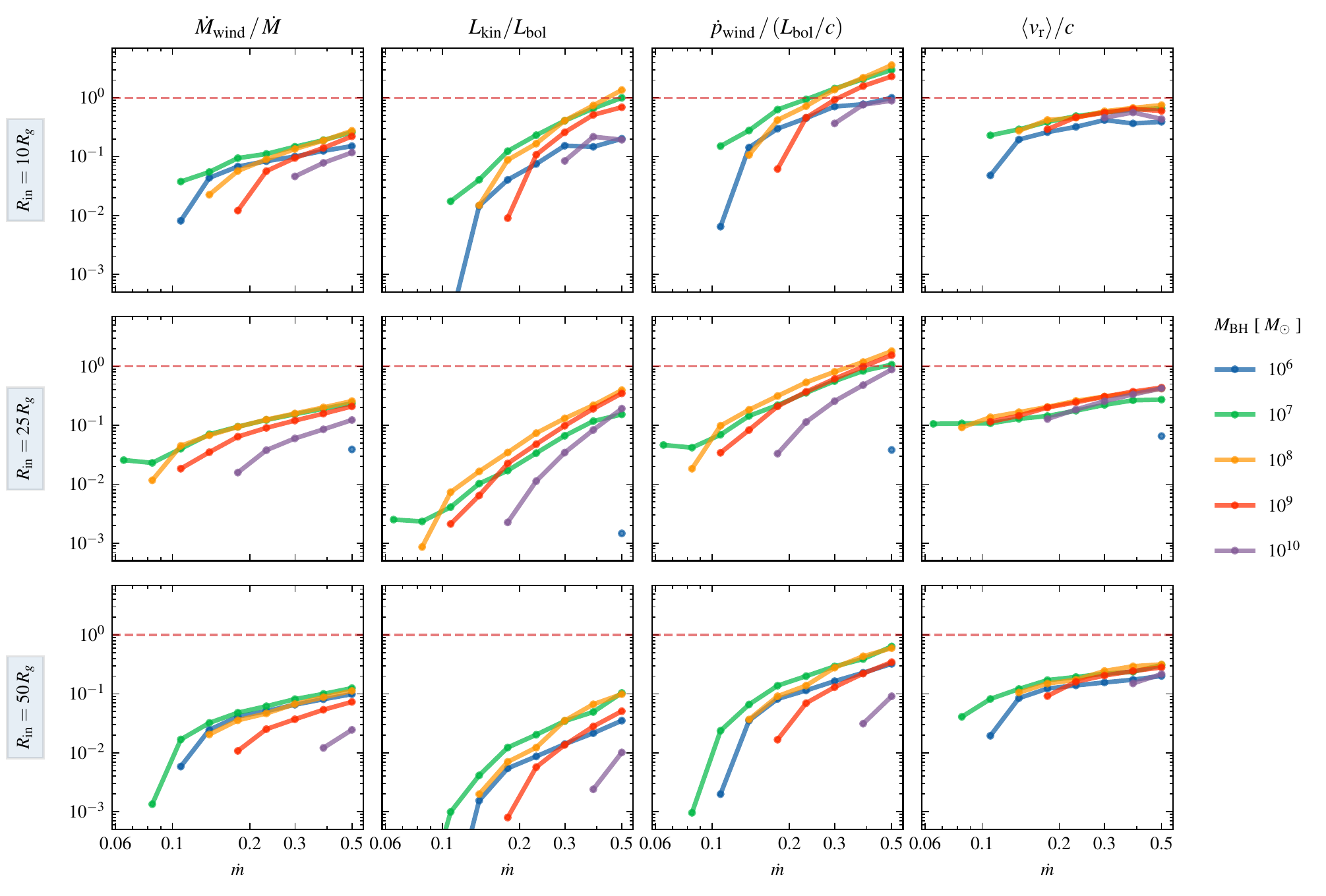}
    \caption{Wind mass loss rate normalised by the mass accretion rate (first panel column), kinetic luminosity normalised by bolometric luminosity (second panel column), momentum loss rate normalised by $L_\mathrm{bol} / c$ (third panel column), and average velocity (fourth panel column) as functions of the Eddington fraction $\dot m$ for different $\M$.The average velocity is taken as \protect $v_\mathrm{r} = \sqrt{2 L_\mathrm{kin} / \dot M_\mathrm{wind}}$.}
    \label{fig:bh_scan}
\end{figure*}

We now investigate the dependence of the wind mass loss rate, kinetic luminosity, momentum loss rate, and average velocity on $\M$, $\dot m$, and $\rin$. We scan the BH parameter range for $\M\in(10^6 - 10^{10})$, $\dot m \in(0.01-0.5)$, and $\rin = (10, 25, 50)\rg$. We fix $f_\mathrm{X} = 0.15$ and recognise that keeping it constant throughout the parameter scan is a limitation, since in reality it depends on $\M$ and $\dot m$.  The results are shown in \autoref{fig:bh_scan}, where all of the quantities have been normalised to their characteristic scales. The red dashed line denotes the limit when quantities become unphysical since the wind is carrying more mass, energy, or momentum than the disc can provide. The first thing to note is that we do not obtain any wind for $\dot m \lesssim 0.06$, regardless of $\M$. This is initially surprising, as the force multiplier is of order $1000$ for cool material, apparently allowing a wind to escape for $\dot{m}>0.001$. However, the initial density drops as $\dot{m}^{1/\alpha}$ (see \autoref{subsection:ic_scaling}) so the X-ray shielding drops dramatically, strongly suppressing the wind. Overall, the weakest winds are seen from the highest ($\M=10^{10}\,\msun$) and lowest ($\M=10^6\,\msun$) black hole masses. This can be explained by the behaviour of the UV fraction (\autoref{fig:uv_fractions}). For the $\M=10^6\,\msun$ case, the UV bright disc annuli are located at large radii, where the disc luminosity is lower; for the $M=10^{10}\msun$ case, $\fuv$ is only high at very small radii, and overall small in the wind launching region. Furthermore, the high disc temperatures expected for the lowest mass systems ($\M = 10^6\,\msun$), especially at high $\dot{m}$, mean that the disc contributes to the ionising X-ray flux. This effect is not considered in our work here, but makes it likely that even our rather weak UV line-driven wind is an overestimate for these systems.

For the values of $\M$ where the wind is robustly generated across the rest of the parameter space, $\M \in (10^7, 10^8, 10^9)\msun$, we find a weak dependence of the normalised wind properties on $\M$. This is expected, since the initial density profile scales with $\M$ and $\dot m$ as (\autoref{subsection:ic_scaling})
\begin{equation}
    n_0 \propto \frac{\dot m^{\frac{1}{\alpha}}}{\M},
\end{equation}
where we ignore the dependence of $\fuv$ on $\M$. The initial wind velocity $v_0 = \vth \propto T^{1/2}$, hence
\begin{equation}
    v_0 \propto \left( \frac{\dot m}{\M} \right)^{1/8} 
\end{equation}
(see \autoref{eq:radiation_flux}.)
This  then implies
\begin{equation}
    \dot M_\mathrm{wind} \propto n_0 \; v_0 \; R^2 \propto \left(\frac{\dot m^{\frac{1}{\alpha}}}{\M}\right) \left(\frac{\dot m}{\M}\right)^{1/8}\left( \M^2\right)
\end{equation}
Since scaling due to the dependence on $v_0$ is particularly weak (it depends on the 1/8-th power), we choose to ignore it so that we can write
\begin{equation}
    \frac{\dot M_\mathrm{wind}}{\dot M_\mathrm{acc}} \propto \frac{\dot m^{\frac{1}{\alpha}} \, \M}{\dot m \M}\propto \dot m^{\frac{1}{\alpha} -1},
\end{equation}
where we have used the fact that $\dot M_\mathrm{acc} = \dot m \dot M_\mathrm{Edd} \propto \dot m \M$. Similarly,
\begin{equation}
    \frac{L_\mathrm{kin}}{L_\mathrm{bol}} \propto \frac{\dot M_\mathrm{wind} v_f^2}{\dot M_\mathrm{acc}} = \dot m^{1 + \frac{1}{\alpha}},
\end{equation}
where $v_f$ is the final wind velocity and we have assumed $v_f \propto \dot m$. This last assumption is verified (for $0.1 \lesssim \dot m \lesssim 0.5$) in the rightmost panel of \autoref{fig:bh_scan}. Consequently, ignoring the scaling of $\fuv$, both the normalised mass loss rate and normalised kinetic luminosity are independent of $\M$. For our value of $\alpha=0.6$, we find $L_\mathrm{kin} \propto \dot m^{2.7} L_\mathrm{bol}$. This scaling is significantly different from the one often assumed in models of AGN feedback used in cosmological simulations of galaxy formation, where the energy injection rate is assumed to be proportional to the mass accretion rate and hence to the bolometric luminosity \citep{schaye_eagle_2015, weinberger_simulating_2017, dave_simba_2019}. However, it is consistent with the results found in the hydrodynamical simulations of \cite{nomura_line-driven_2017} and with current observational constraints \citep{gofford_suzaku_2015, chartas_multiphase_2021}.

For the $\rin = 10 \rg$ case, many parameter configurations of $\M$ and $\dot m$ give rise to winds that are unphysical, since they carry more momentum than the radiation field. This is caused by us not considering the impact that the wind mass loss would have on the disc SED, and an underestimation of the UV opacity in our ray tracing calculation of the UV radiation field, in which we only include the Thomson opacity. We also observe that for the lower and upper ends of our $\M$ range the existence of a wind for different $\dot m$ values depends on the value of $\rin$. This is a consequence of a complex dependence of the failed wind shadow on the initial density and $\rin$.

\begin{figure}
    \centering
    \includegraphics[width=\columnwidth]{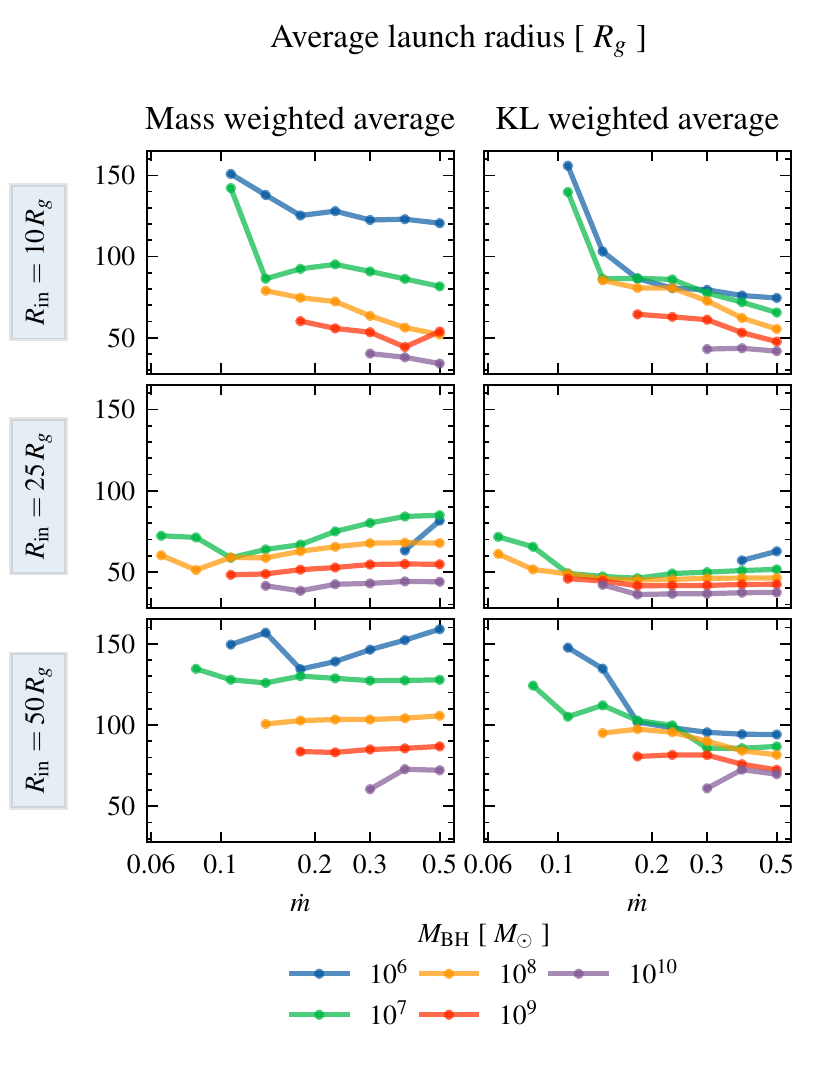}
    \caption{Average launch radius weighted by the trajectories' mass loss rate (left panels) and kinetic luminosity (right panels) as a function of $\M$ and $\dot m$ for the 3 values of $R_\mathrm{in}$.}
    \label{fig:bh_scan_radii}
\end{figure}

We now investigate the geometry of the wind: where it originates and at which angles it flows outwards. Despite the strong dependence of the kinetic luminosity on $\dot m$, the wind launching region does not vary significantly with $\dot m$, as is shown in \autoref{fig:bh_scan_radii}, where we plot the average launch radius, weighted by mass loss rate or kinetic luminosity. The exception is the case $\rin =10\rg$, where the lower inner density at $\rin=10\rg$ (see \autoref{fig:initial_conditions}), and its decrease with $\dot m$ produce a larger failed wind region for $\dot m \lesssim 0.15$. The dependence of the mass weighted average radius with $\M$ is a consequence of the dependence of $\fuv$ on $\M$. Larger $\M$ black holes have the $\fuv$ peak closer in, so the wind carries more mass at smaller radii.

\begin{figure}
    \centering
    \includegraphics[width=\columnwidth]{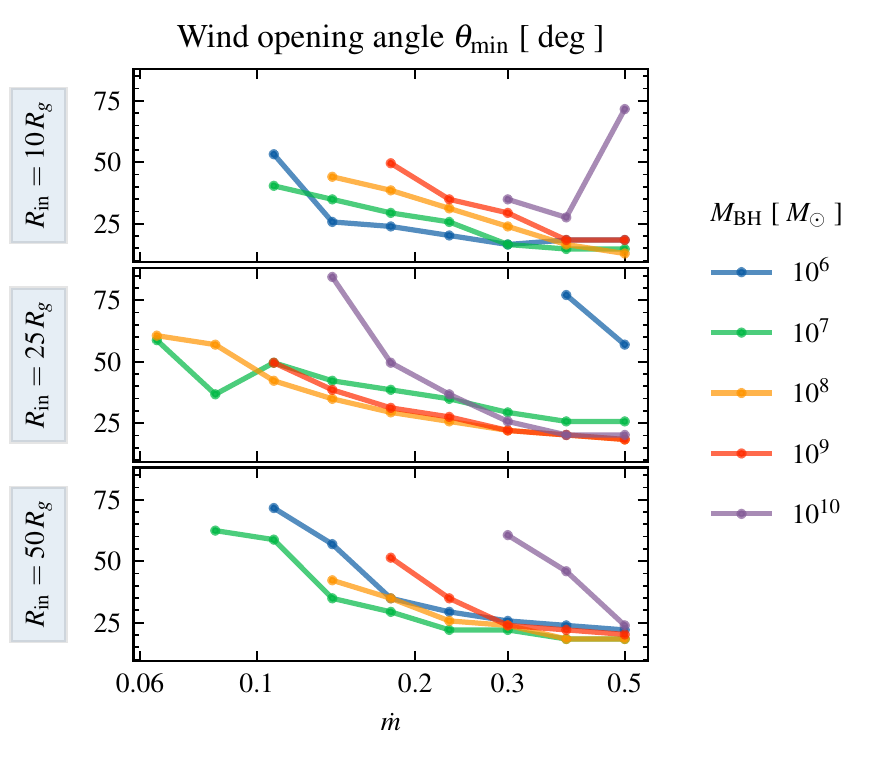}
    \caption{Wind opening angle, measured as the smallest polar angle, $\theta_\mathrm{min}$, of all escaping streamlines as a function of $\M$ and $\dot m$, for the 3 studied $\rin$ values.}
    \label{fig:wind_angle}
\end{figure}

Finally, we plot the wind opening angle for the scanned parameter space in \autoref{fig:wind_angle}. We again find a small dependence on $\M$, except for near the boundaries of our $\M$ range, where the angle is quite sensitive to $\dot m$. The wind flows closer to the equator as we decrease $\dot m$, hence whether the wind is more polar or equatorial depends on the mass accretion rate of the system. This can be explained by considering that lower disc luminosities do not push the wind as strongly from below, and thus the wind escapes flowing closer to the equator.

\subsection{Can UV line-driven winds be UFOs?}

\begin{figure}
    \centering
    \includegraphics[width=\columnwidth]{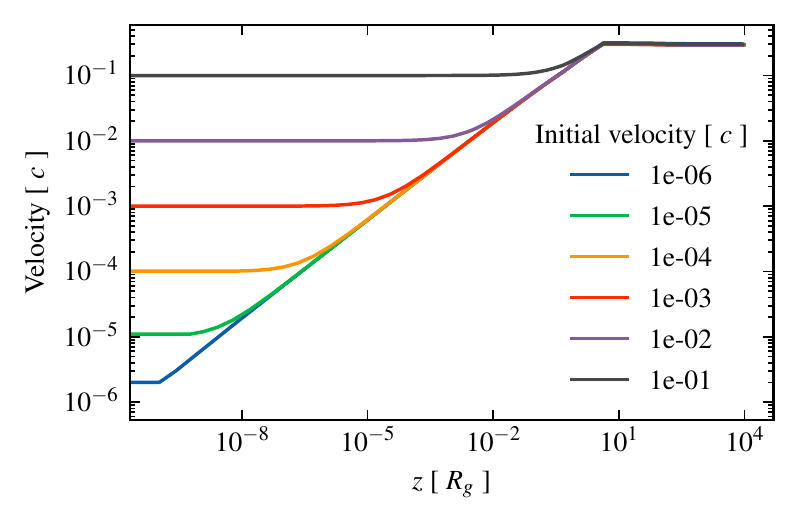}
    \caption{Velocity as a function of $z$ for a trajectory starting at $R=100\rg$ for our fiducial model. Different colours correspond to different initial velocities. }
    \label{fig:velocity_dependence}
\end{figure}

In \qo, wind trajectories could achieve arbitrarily large velocities, often surpassing the speed of light, due to the neglect of relativistic effects. With the introduction of relativistic corrections (\autoref{sec:relativistic}), the wind is always sub-luminal, as we show in \autoref{fig:rin_scan} and \autoref{fig:bh_scan}. Nonetheless, throughout the simulated parameter space, outflows consistently achieve speeds of $(0.1 - 0.8)$ $c$, scaling approximately as $\dot{m}$
with little dependence on 
$\M$. If we limit ourselves to the simulations that conserve the overall momentum and energy of the system, then the simulated wind still achieve speeds in the range $(0.1 - 0.4)$c. This implies that UV line-driving is a feasible mechanism to produce UFOs even when relativistic corrections are included. The final velocity of the wind depends on how much a gas blob can be accelerated while it is shadowed from the X-ray radiation. In \autoref{fig:velocity_dependence}, we plot the velocity profile for a trajectory starting at $R=100\rg$ in our fiducial simulation density grid for different values of the initial velocity. We find that the final velocity of the trajectory is independent of the initial velocity and that the wind is able to drastically accelerate (up to 6 orders of magnitude in velocity) over a very small distance ($\lesssim 1 \rg$), which suggests that line-driving can be very effective even when the X-ray shadowed region is very small.

We thus find that UV line-driving is sufficient to reproduce the range of observed UFO velocities, as opposed to the findings of \cite{luminari_speed_2021}. 
We note that their code assumes an initial density (similar to \qo) rather than calculating it from first principles, and does not include the full ray tracing of both UV and X-rays that are considered here. On the other hand, our treatment of the force multiplier is simplified compared to their calculation, where they use the radiative transfer code XSTAR \citep{kallman_photoionization_2001} to compute the radiation flux absorbed by the wind. Nonetheless, our results show that 
a small shaded region can produce a very fast wind (see 
\autoref{fig:velocity_dependence}), 
and the range of $\rin$ that gives rise to velocities $\geq 0.3 c$ is quite wide (see \autoref{fig:bh_scan_radii}). It then seems
quite likely that UV line-driven winds can indeed reach these velocities and hence be the origin of the majority of UFOs seen. 
There are even higher velocities claimed for a few absorption features in the literature, but these are generally low signal-to-noise detections. 
\section{Conclusions and future work}

In this paper, we have presented \textsc{Qwind3}, a model that builds upon the non-hydrodynamical code \textsc{Qwind} first introduced in \cite{risaliti_non-hydrodynamical_2010} to model UV line-driven winds in the context of AGNs. In this new version, we generalise the CAK formalism for stellar winds to AGNs, which we use to calculate the initial density and velocity with which the wind is launched at the surface of the accretion disc. We have highlighted the importance of correctly accounting for the fraction of luminosity emitted in the UV by each disc annulus, which in other numerical codes is assumed to be a constant over the whole disk. We have also introduced an algorithm to do ray-tracing calculations throughout the wind, allowing us to compute the radiation transfer of the system taking into account the full geometry of the wind and the spatial distribution of the X-ray and UV radiation sources. Furthermore, we have also included special relativistic corrections to the calculation of the radiation flux, which are important for ensuring that the wind does not achieve superluminal speeds. We note two important assumptions that still remain: the simplified dependence of the X-ray opacity on the ionisation parameter and the X-ray luminosity fraction, which we here assume is constant at $f_X=0.15$, independent of black hole mass and mass accretion rate. We will address these issues in the next {\sc Qwind} release. 

We have used the new code to explore under what conditions UV line-driven winds are successfully launched from accretion discs, studying how the normalised mass loss rate, kinetic luminosity, momentum outflow rate, and final velocity change as a function of black hole mass, mass accretion rate, and initial wind launching radius. We find that winds can carry a mass loss rate up to 30\% of the mass accretion rate,
so this can have a moderate impact on the mass accretion rate at radii below the wind launching radius. The next {\sc Qwind} release will address this by reducing the mass accretion rate through the disc, and hence reducing the UV flux produced at small radii, to make the model self-consistent (see also \citealt{nomura_line-driven_2020}).
The current code produces winds where the 
kinetic power and especially the momentum outflow rate can 
be comparable to that of the radiation power at high $\dot{m}$, and even exceed it which is unphysical. This shows that the effect of line blanketing must be important in reducing the UV flux driving the wind below that predicted by electron scattering opacity alone. We will address this in the next {\sc Qwind} release, but 
overall, the wind clearly carries sufficient power to meet the criteria for an efficient feedback mechanism in galaxy formation \citep{hopkins_stellar_2016}.

Our results here show that the outflow velocity is mildly relativistic across a broad parameter space, with velocity $0.1-0.3$~c even excluding the unphysically efficient winds. Thus UV line-driven winds can reach UFO velocities even when special relativistic corrections (radiation drag) are included. This contrasts with \cite{luminari_speed_2021}, who conclude that UV line-driving is not capable of generating such high velocity winds. 
We caution that the two codes make different assumptions about the initial conditions and ray tracing (both of which \textsc{Qwind} does more accurately and self-consistently) as well as the opacity (which their code does better), so it is premature to rule out UV line-driving in favour of other mechanisms such as magnetic driving \citep{blandford_hydromagnetic_1982, fukumura_magnetic_2017} as the origin of UFOs until these factors are all incorporated together. We will explore this more fully in the next \textsc{Qwind} release.

The normalised wind mass loss rate and normalised kinetic luminosity vary substantially as a function of black hole mass and accretion rate, the latter being the most significant factor. The ratio of wind kinetic power to mass accretion rate scales steeply with $\dot{m}$, in contrast to the constant ratio normally assumed in the AGN feedback models currently implemented in cosmological simulations of galaxy formation. Implementing this new more physically-based AGN feedback prescription in simulations will therefore change how galaxies are predicted to evolve across cosmic time.

\section*{Acknowledgements}

AQB acknowledges the support of STFC studentship (ST/P006744/1) and the JSPS London Pre/Postdoctoral Fellowship for Foreign Researchers. CD and CGL acknowledge support from STFC consolidated grant ST/T000244/1. CD acknowledges support for vists to Japan from Kavli Institute for the Physics
and Mathematics of the Universe (IPMU) funding from the National
Science Foundation (No. NSF PHY17-48958).
This work used the DiRAC@Durham facility managed by the Institute for Computational Cosmology on behalf of the STFC DiRAC HPC Facility (www.dirac.ac.uk). The equipment was funded by BEIS capital funding via STFC capital grants ST/K00042X/1, ST/P002293/1, ST/R002371/1 and ST/S002502/1, Durham University and STFC operations grant ST/R000832/1. DiRAC is part of the National e-Infrastructure. This work was supported in part by JSPS Grant-in-Aid for Scientific Research (A) JP21H04488 (KO), Scientific Research (C) JP18K03710 (KO), Early-Career Scientists JP20K14525 (MN). This work was also supported by MEXT as "Program for Promoting Researches on the Supercomputer Fugaku" (Toward a unified view of the universe: from large scale structures to planets, JPMXP1020200109) (KO), and by Joint Institute for Computational Fundamental Science (JICFuS, KO). This paper made use of the Matplotlib \citep{hunter_matplotlib_2007} and SciencePlots \citep{garrett_scienceplots_2021} software packages.




\bibliographystyle{mnras}
\bibliography{bibliography} 




\appendix

\section{Critical point derivation}
\label{app:initial_conditions}

In this appendix we aim to give a detailed and self-contained derivation of the initial conditions for launching the wind from the surface of the accretion disk. Let us first start with the 1D wind equation (now including gas pressure forces),
\begin{equation}
    \label{eq:ic_1d}
    \rho \frac{\dd v_z}{\dd t} = \rho (a_\mathrm{rad}^z - a_\mathrm{grav}^z) - \frac{\partial P}{\partial z}.
\end{equation}
The left-hand side of the equation can be expanded to
\begin{equation}
    \rho \frac{\dd v_z}{\dd t} = \rho \frac{\partial v_z}{\partial t} + \rho v_z \frac{\partial v_z}{\partial z} ,
\end{equation}
where we note that $\partial v_z / \partial t$ = 0, since we are only interested in steady solutions that do not depend explicitly on time. For simplicity, we drop the partial derivative notation going forward, since only derivatives along the $z$ direction are involved. We assume that the wind is isothermal, with an equation of state given by
\begin{equation}
    P = c_s^2 \rho
\end{equation}
where $c_s$ is the isothermal sound speed, which is assumed constant throughout the wind. Using the mass conservation equation (\autoref{eq:mass_conservation}), $\dot M = A\, \rho\, v_z$, where $A = 2\pi r \Delta r$, we can write
\begin{equation}
    \frac{\dd P}{\dd z} =c_s^2\frac{\dd \rho}{\dd z} = -\rho c_s^2 \left(\frac{\dd A / \dd z}{A} + \frac{\dd v_z / \dd z}{v_z}\right),
\end{equation}
so we can rewrite \autoref{eq:ic_1d} as
\begin{equation}
     v_z \frac{\dd v_z}{\dd z}\left(1-\frac{c_s^2}{v_z^2}\right) = a_\mathrm{rad}^z - a_\mathrm{grav}^z + c_s^2\frac{\dd A / \dd z}{A}.
\end{equation}
We now focus on the radiation force term, which is the sum of the electron scattering and line-driving components,
\begin{equation}
    a_\mathrm{rad}^z = a_\mathrm{rad}^{\mathrm{es}, z} + \mathcal M \, a_\mathrm{rad}^{\mathrm{es}, z}.
\end{equation}
When the wind is rapidly accelerating, as it is the case at low heights, the force multiplier is well approximated by its simpler form (see section 2.2.3 of \citetalias{quera-bofarull_qwind_2020}),
\begin{equation}
    \label{eq:fm_simple}
    \mathcal M(t) = k \, t^{-\alpha},
\end{equation}
where $\alpha$ is fixed to $0.6$, $k$ depends on the ionisation level of the gas (in the \citetalias{stevens_x-ray_1990} parameterisation), and $t = \sigmae \, \rho \, \vth \left | \dd v/\dd z \right | ^{-1}$. As discussed in section 3.1.3 of \qo, the thermal velocity, $v_\mathrm{th}$ (\autoref{eq:thermal_velocity}), is computed at a fixed temperature of $T=2.5 \times 10^4$ K. Once again making use of the mass conservation equation we can write
\begin{equation}
    \mathcal M = k\, t^{-\alpha} = \frac{k}{(\sigmae\, \vth\, \rho)^\alpha}\left |\frac{dv}{dz}\right|^\alpha = \frac{k}{\left(\sigmae \vth \, \right)^\alpha}\left(\frac{A}{\dot M} v_z \left|\frac{dv_z}{dz}\right|\right)^\alpha
\end{equation}
and so the full equation to solve with all derivatives explicit is
\begin{equation}
    \begin{split}
        v_z \frac{\dd v_z}{\dd z}\left( 1- \frac{c_s^2}{v_z^2}\right) &= a_\mathrm{rad}^{\mathrm{es}, z} (z) -a_\mathrm{grav}^z (z) \; \\
                                                                      &+ a_\mathrm{rad}^{\mathrm{es}, z}(z)\frac{k}{\left(\sigmae \vth \, \right)^\alpha}\left(\frac{A(z)}{\dot M} v_z \left|\frac{dv_z}{dz}\right|\right)^\alpha \\
                                                                      &+ c_s^2\frac{\dd A(z) / \dd z}{A(z)}.
    \end{split}
\end{equation}
We assume that $\dd v_z / \dd z >0$ always, so the absolute value can be dropped. It is convenient to introduce new variables to simplify this last equation. The first change we introduce is $W=v_z^2/2$, so that $v_z \, \dd v_z / \dd z = \dd W / \dd z$. Next, we aim to make the equation dimensionless, so we consider $R$ as the characteristic length, $B_0 = G\M/R^2$ as the characteristic gravitational acceleration value, $W_0=G\M/R = B_0 R$ as the characteristic $W$ value, $A_0 = 2\pi R \Delta R$ as the characteristic $A$ value, and finally we take the characteristic radiation force value to be 
\begin{equation}
\label{eq:gamma_0}
    \gamma_0 =  \frac{k}{(\sigmae \vth)^\alpha} \; a_\mathrm{rad, 0}^{\mathrm{es}, z},
\end{equation}
where $a_\mathrm{rad, 0}^{\mathrm{es}, z}$ is  \autoref{eq:radiation_acceleration_approx} without considering the $\tauuv$ factor, since we do not include attenuation in this derivation. With all this taken into account, we can write
\begin{equation}
    \label{eq:half_way}
    \begin{split}
        B_0 \frac{dw}{dx} \left(1 - \frac{s}{w}\right) & = -a_\mathrm{grav}^z + a_\mathrm{rad}^{\mathrm{es}, z} \left(1+ \frac{k}{(\sigmae \vth )^\alpha}\left(\frac{A}{\dot M} B_0\frac{dw}{dx}\right)^\alpha\right) \\
                                                       & + c_s^2\frac{\dd A / \dd z}{A},
\end{split}
\end{equation}
where we have defined $x=z/R$, $w=W/W_0$, and $s=c_s^2 / (2 W_0)$. We also define $a=A/A_0$, and $\varepsilon = \dot M / \dot M_0$ with
\begin{equation}
    \dot M_0 = \alpha (1-\alpha)^{(1-\alpha) / \alpha} \frac{(\gamma_0 A_0)^{1 / \alpha}}{(B_0 A_0)^{(1-\alpha) / \alpha}}.
    \label{eq:Mdot0_def}
\end{equation}
This last definition may seem a bit arbitrary, but it is taken such that $\varepsilon = 1$ corresponds to the classical \citetalias{castor_radiation-driven_1975} $\dot M$ value for O-stars. \autoref{eq:half_way} then becomes
\begin{equation}
    \begin{split}
        \frac{\dd w}{\dd x} \left(1-\frac{s}{w}\right) &= \frac{(-a_\mathrm{grav}^z + a_\mathrm{rad}^{\mathrm{es}, z})}{B_0} + \frac{1}{\alpha^\alpha(1-\alpha)^{1-\alpha}}\frac{a_\mathrm{rad}^{\mathrm{es}, z}}{a_\mathrm{rad,0}^{\mathrm{es}, z}} \left(\frac{a}{\varepsilon}\frac{\dd w}{\dd x}\right)^\alpha \\
    &+ \frac{4sx}{a},
    \end{split}
\end{equation}
where we have used
\begin{equation}
    a = \frac{A}{A_0} = \frac{2\pi r \Delta r}{2\pi r_0 \Delta r_0} = \frac{r^2}{r_0^2},
\end{equation}
so that
\begin{equation}
    \frac{c_s^2}{B_0}\frac{\dd A / \dd z}{A} = \frac{4sx}{a}.
\end{equation}
We note that we have assumed that the area $A$ changes with $z$ like the 2D solution, despite the fact that we are considering here a 1D wind. This small correction guarantees that we find critical-point like solutions for all initial radii, since it guarantees that the \citetalias{castor_radiation-driven_1975} conditions for the existence of a critical point are satisfied.
Finally, we define
\begin{equation}
\label{eq:nozzle_f}
    f = \frac{1}{\alpha^\alpha (1-\alpha)^{1-\alpha}}\frac{a_\mathrm{rad}^{\mathrm{es}, z}}{a_\mathrm{rad, 0}^{\mathrm{es}, z}},
\end{equation}
and 
\begin{equation}
\label{eq:nozzle_h}
    h = \frac{(a_\mathrm{grav}^z - a_\mathrm{rad}^{\mathrm{es}, z})}{B_0} - 4sxa,
\end{equation}
so that
\begin{equation}
    \frac{\dd w}{\dd x} \left(1-\frac{s}{w}\right) = -h(x) + f(x) \left(\frac{a}{\varepsilon}\frac{\dd w}{\dd x}\right)^\alpha,
\end{equation}
which is the dimensionless wind equation. The choice of sign in $h(x)$ is to make the further steps clearer.
It is useful to interpret this equation as an algebraic equation for $w'=\dd w / \dd x$,
\begin{equation}
    F(x,w,w') =  w' \left(1-\frac{s}{w}\right) + h(x) - f(x) \left(\frac{a}{\varepsilon}w'\right)^\alpha = 0.
    \label{eq:CAK_F_wprime}
\end{equation}
This has the same general form as Equation~26 in \citetalias{castor_radiation-driven_1975}, which applies to a spherical wind from a star, but the functions $h(x)$ and $f(x)$ are different. We note that $f(x)>0$ and $0<\alpha<1$, but $h(x)$ can have either sign. (Also note that we have chosen the opposite sign for $h(x)$ to CAK, for later convenience.)
Given values of $x$ and $\varepsilon$, we can distinguish 5 different regions of solutions for $w'$ (we follow the original enumeration of regions by \citetalias{castor_radiation-driven_1975}):

\begin{itemize}
    \item Subsonic stage ($w<s$): We distinguish two cases
        \begin{itemize}
            \item Region V: If $h(x)<0$, then all the terms in Equation~\ref{eq:CAK_F_wprime} are negative and there is no solution for $w'$.
            \item Region I: If $h(x)>0$, $F(w'=0)$ is positive and as $w'$ increases, $F$ goes to negative values crossing $F=0$ once, so there is one solution for $w'$.
        \end{itemize}
    \item Supersonic stage ($w >s$): We have three regions
        \begin{itemize}
            \item Region III: If $h(x)<0$, $F(w'=0)$ is negative and as $w'$ increases, $F$ goes to positive values crossing $F=0$ once, so there is one solution for $w'$.
            \item If $h(x) > 0$, $F(w'=0)$ is positive, then $F$ initially decreases until its minimum, $w_\mathrm{min}'$, and then increases again. If $F(w_\mathrm{min}') >0$, we have no solution  for $w'$ and if $F(w_\mathrm{min}') \leq 0$ we have two solutions, which are the same if $F(w_\mathrm{min}') = 0$.
                The minimum can be found by solving
                \begin{equation}
                    \frac{\partial F}{\partial w'} = 0,
                \end{equation}
                which gives
                \begin{equation}
                \label{eq:w_min}
                    w_\mathrm{min}' = \left( \frac{1-s/w}{\alpha f (a/\varepsilon)^\alpha}\right)^\frac{1}{\alpha -1},
                \end{equation}
                so we have
                \begin{itemize}
                    \item Region IV: If $F(w_\mathrm{min}') > 0$, there is no solution.
                    \item Region II: If $F(w_\mathrm{min}') \leq 0$, there are two solutions, one with $w' < w_\mathrm{min}'$ and the other with $w' > w_\mathrm{min}'$
                \end{itemize}
        \end{itemize}
\end{itemize}

We assume that the wind starts subsonic ($w<s$), so it has to start in Region~I, since that is the only subsonic region with a solution. We note that $h(x\to\infty) < 0$, since for large $x$ both the gravitational and radiation force are small. Assuming that the wind ends as supersonic (w>s) and extends to $x\to\infty$, this means that the wind must end at Region~III. However, because $h(x)>0$ in Region~I and $h(x)<0$ in Region~III, these two regions must be connected by Region~II in between. At the boundary between Regions~I and II, $w=s$ and $h(x)>0$, while at the boundary between Regions~II and III, $w>s$ and $h(x)=0$.

Considering first the boundary between Regions~I and II, setting $w=s$ in \autoref{eq:CAK_F_wprime} gives
\begin{equation}
    h - f \left(\frac{a}{\varepsilon} w'\right)^\alpha = 0,
\end{equation}
with $h>0$. Considering \autoref{eq:w_min} for $w_\mathrm{min}'$ as $w \to s^{+}$, we find that the wind solution at this boundary must lie on the branch with $w' < w_\mathrm{min}'$.

Considering next the boundary between Regions~II and III, setting $h=0$ in \autoref{eq:CAK_F_wprime} (neglecting the $w'=0$ solution) gives
\begin{equation}
    1-\frac{s}{w} - f \left(\frac{a}{\varepsilon}\right)^\alpha w'^{\alpha-1} = 0,
\end{equation}
Again using \autoref{eq:w_min} and $w>s$, we find that the wind solution at this boundary must lie on the branch with $w' > w_\mathrm{min}'$.

However, we assume that $w'$ must be continuous throughout the wind, which would not be the case if the two branches of region II were not connected. Therefore, both branches of region II must coincide at some point, so the condition
\begin{equation}
    F(w_\mathrm{min}') = 0
\end{equation}
must hold at that point, with $w' = w_\mathrm{min}'$ for both branches. This point is in fact the critical point of the solution, and $\partial F/\partial w' = 0$ there. Upon substitution of $w_\mathrm{min}'$, this condition is equivalent to
\begin{equation}
    \label{eq:nozzle_function_sonic}
    \varepsilon \, \left(1 - \frac{s}{w} \right) = \alpha (1-\alpha)^\frac{1-\alpha}{\alpha} \, \frac{f^{1/\alpha} a}{h^\frac{1-\alpha}{\alpha}}.
\end{equation}
The right-hand side of the equation is usually referred to as the Nozzle function $\mathcal N$,
\begin{equation}
\label{eq:nozzle_function}
    \mathcal N(x) = \alpha (1-\alpha)^\frac{1-\alpha}{\alpha} \frac{f^{1/\alpha}a}{h^\frac{1-\alpha}{\alpha}}.
\end{equation}
We define $x=x_c$ to be the position of the critical point. We now assume that at the critical point the wind is highly supersonic ($w \gg s$). We verify this assumption in \autoref{subsection:verify_critical_point}. Then $1-s/w \approx 1$, and so the normalised mass loss rate is given by
\begin{equation}
    \varepsilon = \mathcal N(x_c).
    \label{eq:eps_n_xc}
\end{equation}

We now show that the location of the critical point $x_c$ is at the minimum of $\mathcal N(x)$. Let us consider the total derivative of $F$ with respect to $x$ taken along a wind solution. Since $F=0$ at all points along the solution,
\begin{equation}
    \frac{\dd F}{\dd x} = \frac{\partial F}{\partial x} + \frac{\partial F}{\partial w}w' + \frac{\partial F}{\partial w'}\frac{\dd w'}{\dd x} = 0.
\end{equation}
Because of our assumption $w\gg s$, we have $\frac{\partial F}{\partial w} = 0$, and, at the critical point, $\frac{\partial F}{\partial w'} = 0$, so that
\begin{equation}
    \left .\frac{\dd F}{\dd x}\right |_{w'_\mathrm{min}, x_c} = \left .\frac{\partial F}{\partial x}\right |_{w'_\mathrm{min}, x_c} = 0,
\end{equation}
which gives (a prime denotes $\dd / \dd x$)
\begin{equation}
\label{eq:dF_dx}
    h' - \left(\frac{w'_\mathrm{min}}{\varepsilon}\right)^\alpha 
    \frac{\dd}{\dd x} \left( f a^{\alpha} \right) = 0.
\end{equation}
Now using \autoref{eq:w_min} for $w'_\mathrm{min}$ along with \autoref{eq:eps_n_xc} and \autoref{eq:nozzle_function}, we obtain $w'_\mathrm{min} = \frac{\alpha}{(1-\alpha)} h$. Substituting in \autoref{eq:dF_dx} and using \autoref{eq:eps_n_xc} and \autoref{eq:nozzle_function} again, we obtain 
\begin{equation}
(1-\alpha) \frac{h'}{h} -  \frac{1}{f a^{\alpha}} \frac{\dd}{\dd x} \left( f a^{\alpha} \right) = 0.
\end{equation}
Comparing to \autoref{eq:nozzle_function}, we see that this is equivalent to the condition $\dd \mathcal{N}/\dd x =0$. Therefore the critical point happens at an extremum of $\mathcal{N}(x)$, but since the nozzle function is not upper bounded, the extremum has to be a minimum. We have therefore shown that the critical point $x_c$ corresponds to the minimum of the nozzle function $\mathcal{N}(x)$.

\begin{figure}
    \centering
    \includegraphics[width=\columnwidth]{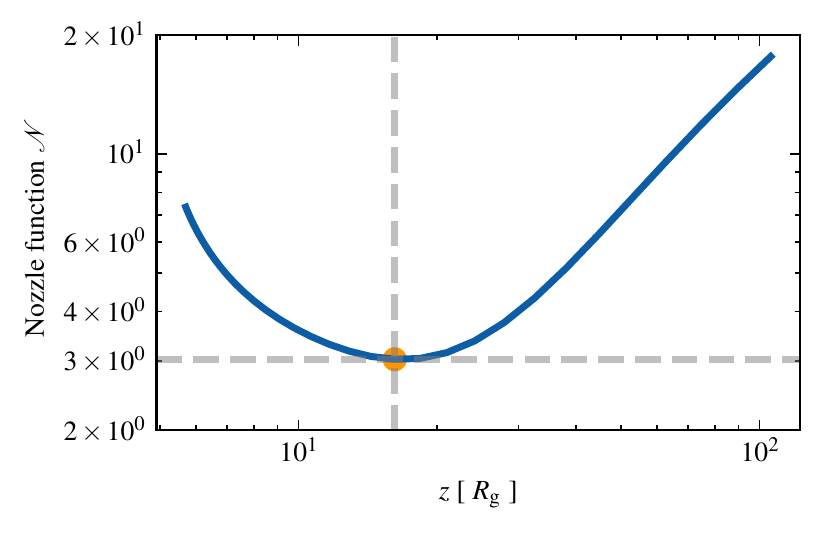}
    \caption{Nozzle function for $\protect\M=10^8\msun$, $\protect\dot m =0.5$, $k=0.03$, and $R=20 \rg$}  
    \label{fig:nozzle_function}
\end{figure}

In \autoref{fig:nozzle_function}, we plot the nozzle function for $\M = 10^8\msun$, $\dot m =0.5$, $k=0.03$, and $R=20\rg$ and we highlight the position of the critical point. The value of the nozzle function at the critical point determines the mass-loss rate along the streamline, through $\dot M = \varepsilon \dot M_0$, so that the surface mass loss rate is given by
\begin{equation}
    \dot \Sigma = \frac{\dot M}{A}.
\end{equation}

{}

It is interesting to point out that the position of the critical point and the value of $\epsilon$ do not depend on the chosen value of $k$ in the force multiplier parametrisation (\autoref{eq:fm_simple}), however, 
the mass loss rate does depend on $k$ through the value of $\dot{M}_0$.
In the \citetalias{stevens_x-ray_1990} parametrisation, $k$ is a function of the ionisation state of the gas, $k=k(\xi)$, so the mass loss rate directly depends on the ionisation conditions at the critical point location. Since this would make our results dependent on the modelling of the vertical structure of the disc, which is out of scope for the purpose of this work, we assume that the gas is always ionised when it reaches the critical point, so we take $k=0.03$, corresponding to the minimum value of $k$ in \citetalias{stevens_x-ray_1990}. Similarly, we set the initial height of the wind to $z=0$ to avoid dependencies on the disc vertical structure.

\subsection{Scaling of the initial conditions with BH properties}
\label{subsection:ic_scaling}

\begin{figure}
    \centering
    \includegraphics{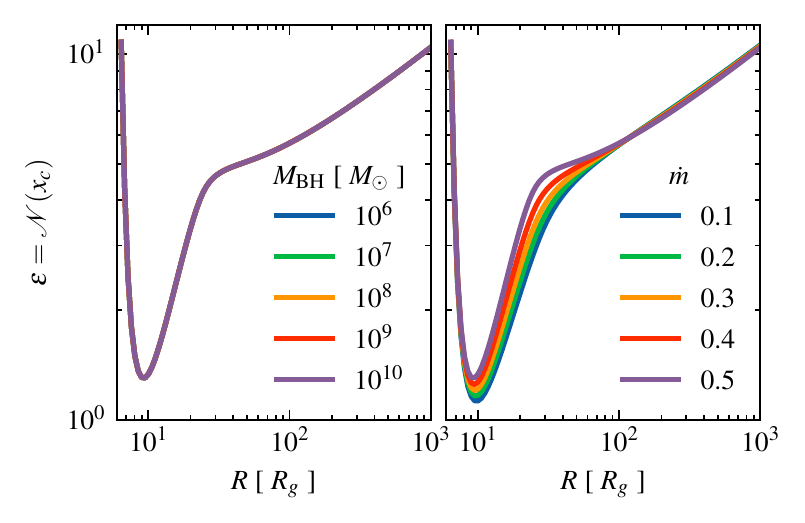}
    \caption{Value of the nozzle function at the critical point ($\varepsilon = \mathcal N(x_c)$) as a function of radius for varying $\M$ (left panel) with $\dot m =0.5$ and varying $\dot m$ (right panel) with $\M=10^8 \msun$}
    \label{fig:nozzle_scaling}
\end{figure}

Since we explore the BH parameter space in \autoref{sec:results_bh}, it is useful to assess how the initial number density and velocity of the wind scale with $\M$ and $\dot m$. To this end, we need to determine how the values of $\varepsilon$ and $\dot \Sigma_0$ change with $\M$ and $\dot m$. For this analysis, we ignore the dependence of $\fuv(R)$ on $M$ and $\dot m$. We also have $\vth(R) \propto T^{1/2} \propto (\M \dot{M}/R^3)^{1/8}$ (using \autoref{eq:radiation_flux}). Accounting for the fact that the disk size scales as $R \propto \rg \propto \M$, this gives $v_0 = \vth \propto (\dot{m}/\M)^{1/8}$, where $v_0$ is the initial velocity. This is a weak dependence, so we ignore it here. Looking at \autoref{eq:gamma_0}, and using \autoref{eq:radiation_acceleration_approx}, we get $\gamma_0 \propto \M \dot{M}/R^3 \propto \dot m / \M$. Similarly, $B_0 \propto \M/R^2 \propto 1 / \M$. Using \autoref{eq:Mdot0_def}, we then have $\dot \Sigma_0 \propto \dot{M}_0/A_0 \propto \gamma_0^{1/\alpha}/B_0^{(1-\alpha)/\alpha} \propto \dot m^{1/\alpha} / \M$.

The scaling of $\varepsilon = \mathcal{N}(x_c)$ is a bit more complicated, since it depends on the exact position of the critical point for each value of $\M$, $\dot m$, and $R$. In \autoref{fig:nozzle_scaling}, we plot the values of $\varepsilon$ as a function of radius for varying $\M$ (left panel) and $\dot m$ (right panel), ignoring the dependence of $\fuv$ with $\M$ and $\dot m$ (we set $\fuv = 1$). We note that $\varepsilon$ does not scale with $\M$, and changes very little with $\dot m$. Including the dependence of $\fuv$ with $\M$ and $\dot m$, effectively reduces the value of $\varepsilon$ at the radii where $\fuv$ is small, but it does not change substantially in the radii that we would expect to launch an escaping wind. We can then conclude that $\varepsilon$ has a very weak scaling with $\M$ and $\dot m$ so that
\begin{equation}
    \label{eq:sigma_scaling}
    \dot\Sigma \propto \dot\Sigma_0 \propto \frac{\dot m^{\frac{1}{\alpha}}}{\M},
\end{equation}
which is the same result that \cite{pereyra_steady_2006} found in applying the \citetalias{castor_radiation-driven_1975} formalism to cataclysmic variables.

\subsection{Verification of the critical point conditions}
\label{subsection:verify_critical_point}

\begin{figure}
    \centering
    \includegraphics[width=\columnwidth]{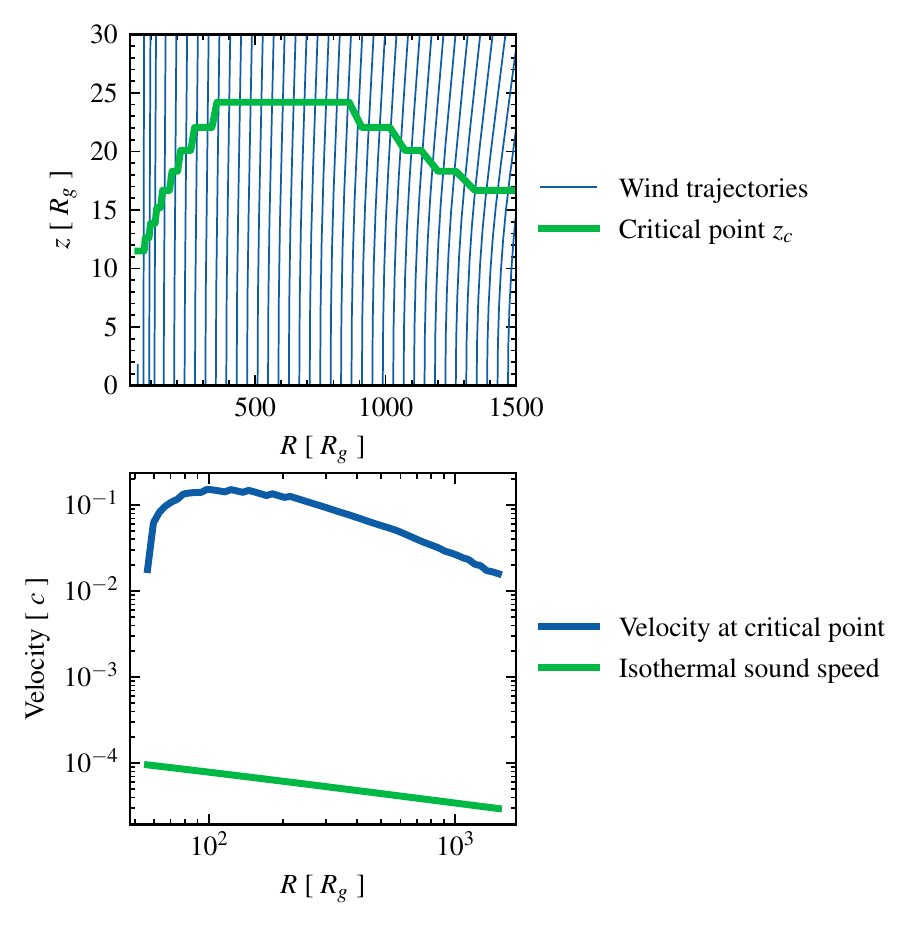}
    \caption{Results for fiducial case $\protect\M=10^8\msun$ and $\protect\dot m =0.5$. Top panel: wind trajectories compared to the critical point position, plotted on a linear scale. Bottom panel: Velocity at the critical point plotted on a log scale}
    \label{fig:verify_critical_point}
\end{figure}

For our fiducial case (\autoref{sec:results}), we plot the critical point location compared to the wind trajectories in \autoref{fig:verify_critical_point}. All escaping trajectories are vertical at the critical point, so our treatment of the wind as a 1D flow for the initial conditions derivation is justified. Nonetheless, we emphasise again that this treatment does not hold for the inner failed wind. The wind is highly supersonic (~$10^3$ times the sound speed) at the critical point, as shown in the bottom panel of \autoref{fig:verify_critical_point}, so our assumption $w\gg s$ is validated.

\section{Intersection of trajectories}
\label{app:intersections}

Once we have solved the equations of motion of the different gas blobs, it is common for the resulting trajectories to cross each other. Trajectories of gas elements computed using our ballistic model should not be confused with the streamlines of the actual wind fluid, since the latter cannot cross each other as it would imply the presence of singular points where the density and the velocity fields are not well defined.

Nonetheless, if we aim to construct a density and velocity field of the wind, we need to define the density and velocity at the crossing points. To circumvent this, once two trajectories intersect, we terminate the one that has the lowest momentum density at the intersection point.

To determine at which, if any, point two trajectories cross, we consider a trajectory as a collection of line segments $\{s_i\}$. Two trajectories $\{s_i\}$, and $\{t_j\}$ cross each other if it exists $i, j$ such that $s_i \cap t_j \neq 0$.

Suppose the line segment $s_i$ is bounded by the points $\bmath{p_1}$ and $\bmath{p_2}$ such that $\bmath{p_2} = \bmath{p_1} + \alpha' (\bmath{p_2} - \bmath{p_1})$ with $\alpha' \in [0,1]$. Similarly, $t_j$ is limited by $\bmath{q_1}$ and $\bmath{q_2}$ such that $\bmath{q_2} = \beta' (\bmath{q_2} - \bmath{q_1})$ with $\beta' \in [0,1]$. The condition that $s_i$ intersects $t_j$ is equivalent to finding $\alpha$, $\beta$ $\in [0,1] \times [0,1]$ such that
\begin{equation}
    \bmath{p_1} + \alpha' (\bmath{p_2} - \bmath{p_1}) = \bmath{q_1} + \beta' (\bmath{q_2} - \bmath{q_1}),
\end{equation}
which corresponds to the linear system $\mathbfss{A}\bmath{x} = \bmath{b}$ with
\begin{equation}
    \mathbfss{A} = \left(\bmath{p_2}-\bmath{p_1}, \bmath{q_1}-\bmath{q_2}\right),
\end{equation}
$\bmath{x} = \left(\alpha', \beta'\right)^\intercal$, and $\bmath{b} = \bmath{q_1} - \bmath{p_1}$. The segments intersect if this linear system is determined with solution inside the unit square.


\bsp	
\label{lastpage}
\end{document}